\documentclass{aa}  
\usepackage{aalongtable,lscape}
\usepackage{graphicx}
\usepackage[figuresright]{rotating}
\usepackage{natbib}
\usepackage{subfigure}
\usepackage{caption}
\usepackage{txfonts}
\newcommand{\nodata}{\mbox{$...$}} 
\begin{document}
   \title{The host in blue compact galaxies:}

   \subtitle{Structural properties and scaling relations}

   \author{Ricardo Amor\'in \inst{1}\inst{2}, J. Alfonso L. Aguerri \inst{1}, 
     Casiana Mu\~noz-Tu\~n\'on \inst{1},
          \and
           Luz M. Cair\'os \inst{3}
          }

    \offprints{R.O. Amor\'in}

   \institute{Instituto de Astrof\'isica de Canarias (IAC), 
     V\'ia L\'actea S/N, E-38200 La Laguna, Tenerife, Spain\\
          \email{ricardo.amorin@iac.es}, 
            \email{casiana@iac.es}, 
	    \email{jalfonso@iac.es}
\and
     Instituto de Astrof\'isica de Andaluc\'ia (IAA), 
     Camino Bajo de Huetor, 50, 18003 Granada, Spain \\
     \email{amorin@iaa.es}
\and
             Astrophysikalisches Institut Potsdam, An der Sternwarte 16, 
     D-14482 Potsdam, Germany\\ 
             \email{luzma@aip.de}
                       }

   \date{Received ..., ...; accepted ..., ...}

   \abstract
   {}
   {We characterise the underlying stellar host in a sample of 20 blue compact galaxies 
(BCGs), by fitting their two-dimensional light distributions. 
 Their derived host structural parameters and those of eight other BCGs 
already obtained in a previous paper are related to galaxy properties, such as colours and gas content. 
These properties are also compared with those of other galaxy types.}
   {The structural parameters of the host were derived by fitting a two-dimensional PSF-convolved 
S\'ersic model to deep optical images in several bandpasses ($B$, $V$, $R$, $I$). We followed a 
fitting technique that consists in the accurate masking-out of the star-forming regions in several steps.}
   {All the BCG hosts but one show low S\'ersic indexes ($0.5\la n \la 2$), with mean 
effective radius $<$$r_{\rm e,B}$$> =$ 1.11$\pm$0.74 kpc and mean surface brightness 
$<$$\mu_{\rm e,B}$$> =$ 22.59$\pm$0.68 mag arcsec$^{-2}$. Host effective radii scale linearly 
with their luminosity, while $n$ and $\mu_{\rm e}$ do not. In addition, host colours and structural 
parameters are not linearly correlated. Overall, the flux enhancement caused by the starburst is 
about 0.8 mag, while their $B-R$ colours decrease by about 0.2 mag. Galaxies with more luminous and 
extended hosts show larger and luminous starburst components, whereas the relative strength of 
the burst ($L_{\rm burst}/L_{\rm host}$) does not show any significant dependence on the host luminosity 
(or mass). While hosts show $B-R =$ 0.95$\pm$0.26 in median, galaxies with redder hosts ($<$$B-R$$> 
=$ 1.29$\pm$0.10) and with bluer hosts ($<$$B-R$$> =$ 0.66$\pm$0.10) are distinguished among the more 
and less luminous systems, respectively. Overall, BCG hosts are more compact (by a factor $\sim$2) 
and have higher central surface brightnesses (by about $\sim$2 mag) than dIs and most dEs. BCG 
hosts and isolated dIs are indistinguishable in the $B$-band Tully--Fisher relation (TFR). We found 
that about 50--60\% of the galaxies are more underluminous than those late-type discs with the 
same circular velocity. This feature is more important when luminosities are converted into stellar 
masses, while it tends to diminish when the H{\sc i} gas mass is added. Deviations among host masses 
for a given circular velocity from the stellar TFR correlate with their H{\sc i} mass-to-luminosity 
ratio ($M_{\rm HI}/L_{\rm B}$), whereas deviations from the gas$+$stellar TFR do not. Overall, our 
findings suggest that the baryonic mass in BCGs tends to normal values, but BCGs tend to be 
inefficient by producing stars, especially toward the low-mass, gas-rich, and bluest hosts, in a 
similar way to dIs. 
}
   {}


   \maketitle
%
\section{Introduction}
Blue compact galaxies \citep[BCGs,][]{Zwicky65} are gas-rich objects that are 
currently experiencing a strong burst of star formation \citep{C01a,GdP03}. Their 
main characteristics were summarised by \citet{ThuanMartin81} as having low 
luminosity (those BCGs less luminous than $M_{\rm B}\ga-$18 are commonly referred 
to as ``blue compact dwarfs''), compactness, and emission line spectra, similar 
to the H{\sc ii} regions in spiral galaxies. Moreover, they tend to be low-metallicity 
systems \citep[$Z_{\odot}/50 \la Z \la Z_{\odot}/2$, e.g.][see also the review by 
Kunth \& {\"O}stlin, 2000]{Terle91} and present 
high star formation rates \citep[0.1--10 M$_{\odot}$ yr$^{-1}$, e.g.][]{Fanelli88,GdP03}.  
Initially these two combined properties led to conjecture that BCGs are pristine galaxies, 
in the process of forming their first generation of stars \citep{sar70}.  
However, several deep photometric studies in the optical \citep[e.g.,][ hereafter
 C01a,b]{LT86,Kunth88,Telles,Papade96a,Doublier1,Doublier2,C00,ByO02,C01a,C01b} and in the 
near-infrared \citep[e.g.,][]{James94,Doublier3,Noeske03,Noeske05,C03} have demonstrated 
that virtually all BCGs have an older underlying stellar ``host'' in addition to the present
 starburst. This supports the idea that BCGs are not truly primordial galaxies, but 
instead older systems undergoing transient periods of strong star formation \citep{mas99}.
 The stellar host generally extends several kpc from the usually centrally concentrated 
star-forming regions. Thus, the host is generally detectable only at low surface brightness levels,
 showing elliptical isophotes, and displaying the red colours indicative of an old stellar
 population \citep[Papaderos et al., 1996b, hereafter P96b; C01a,b;][]{C02,C03,ByO02}. 

Despite many studies over several years of investigation, certain questions  concerning BCGs still 
remain open. The assessment of the structural properties of their host 
galaxy is one of the most important issues in any research of dwarf galaxies. 
Deriving the ages and chemical abundances of the host is essential to establishing the 
evolutionary state and the star formation history of the BCG class.
Furthermore, the possibility of evolutionary connections linking BCGs and other dwarf 
galaxies, namely dwarf irregulars (dIs), dwarf elliptical (dEs), and low surface brightness 
(LSB) galaxies, is a fundamental issue that has still not benn resolved. In particular, the comparison 
of the structural properties, average colours, and colour gradients between the different 
dwarf classes are crucial to testing the proposed evolutionary scenarios and elaborating a general 
view of their formation and evolution \citep[]{Thuan85,DyPh88,Papade96a,Mar97,Mar99,Caon05}. 

For carrying out quantitative studies of the host's properties and go further into the above 
problems, the main handicap is the host faintness (typically $\mu_{\rm B}$$\ga$23 mag arcsec$^{-2}$), 
which requires a great deal of observational and analysis effort. Moreover, the contamination
 caused by the starburst makes the derivation of accurate structural parameters and colours of
 the host still more complicated.
In the past decade, there have been several studies devoted to deriving the structural parameters 
of the host, in the optical and in the NIR, by analysing their radial surface brightness profiles 
\citep[][C01a,b]{Doublier1,Doublier2,Mar97,Noeske03,Noeske05,Vadu05,GdP03,Caon05,C03,C02}. 
However, the derived structural parameters of the host component have been very dependent on 
several factors, such as the extraction of the radial profile and the fitting procedure. Each of 
the different methods has its own drawbacks, resulting in information loss from the image 
\citep{Baggett}. Moreover, the strong dependence on the quality of the analysed dataset, the model 
used to describe the host light profile, and the accuracy in subtracting , i.e.\ masking out, the 
starburst emission from the BCG light profile, have in general provided discrepant results
 \citep[see][]{Kunth&Ostlin,GdP03,C00,C03,Caon05} which could lead to unreliable conclusions. 
 
Two-dimensional algorithms that able to derive structural parameters directly from the galaxy image, 
e.g.\ {\sc gim2d} \citep{Simard98}, {\sc galfit} \citep{Peng02}, {\sc gasp-2d} \citep{Jairo07} have 
opened a new perspective on the  BCG hosts. In \citet{PaperI}, hereafter Paper~I, 
we developed a new 2D technique based on {\sc galfit} for fitting the host in BCGs. 
Their advantages and limitations were discussed at some length by using simulated galaxies and 
observations of eight carefully selected BCGs for which a deep 1D analysis had previously been 
carried out by \citet{Caon05}. One of the main conclusions of Paper~I is that if using accurate masks 
and performing consistency checks, the 2D fitting method is robust and allows us to extract accurate 
structural parameters. An important strength of the technique is the isolation of the starburst emission. 
Accurate masks closely following the starburst dominant region in shape and size allow us to fit the 
host in a larger portion of the galaxy and in a wider range of surface brightnesses. 

In this paper we present host 2D fits for a large sample of BCGs by using the same technique 
as in Paper~I. The goal of this paper is to derive the structural parameters of 20 BCG for which 
underlying emission or low surface brightness features have been detected.
We also included the 8 BCGs previously fitted in Paper~I, finishing the 
2D characterisation of the large sample, representative of the BCG class, already studied in 
C01a by using a 1D technique. 
Furthermore, we aim to explore the global properties of the hosts derived from their 
structural parameters, colours and H{\sc i} gas content, as well as their position in scaling relations, 
such as the Tully--Fisher relation, compared with those of other galaxy types.         
The paper is structured as follows. In \S~2 we present the data. In \S~3 we summarise the main aspects 
of the methodology. Our results are presented in \S~4 and discussed in \S~5. Finally, \S~6 summarises 
our main conclusions.      

\section{Sample}
 
The full sample is composed of 28 star-forming galaxies, all of them already studied by 
our group in several previous papers \citep[e.g., C01a,b,][Paper~I]{Caon05}. 
A complete description of the sample,
\footnote{see http://www.iac.es/proyect/GEFE/BCDs/BCDframe.html} observations, data reduction, 
calibration procedures, and surface photometry, as well as a collection of colour maps and 
H$\alpha$ images, were presented in C01a and C01b. 
The selection criteria, described in C01a, include low-luminosity, 
compactness, and spectra similar to those of H{\sc ii} regions, when available. 
   \begin{figure*}
     \centering
    \resizebox{\hsize}{!}{\includegraphics[angle=0,width=15cm]{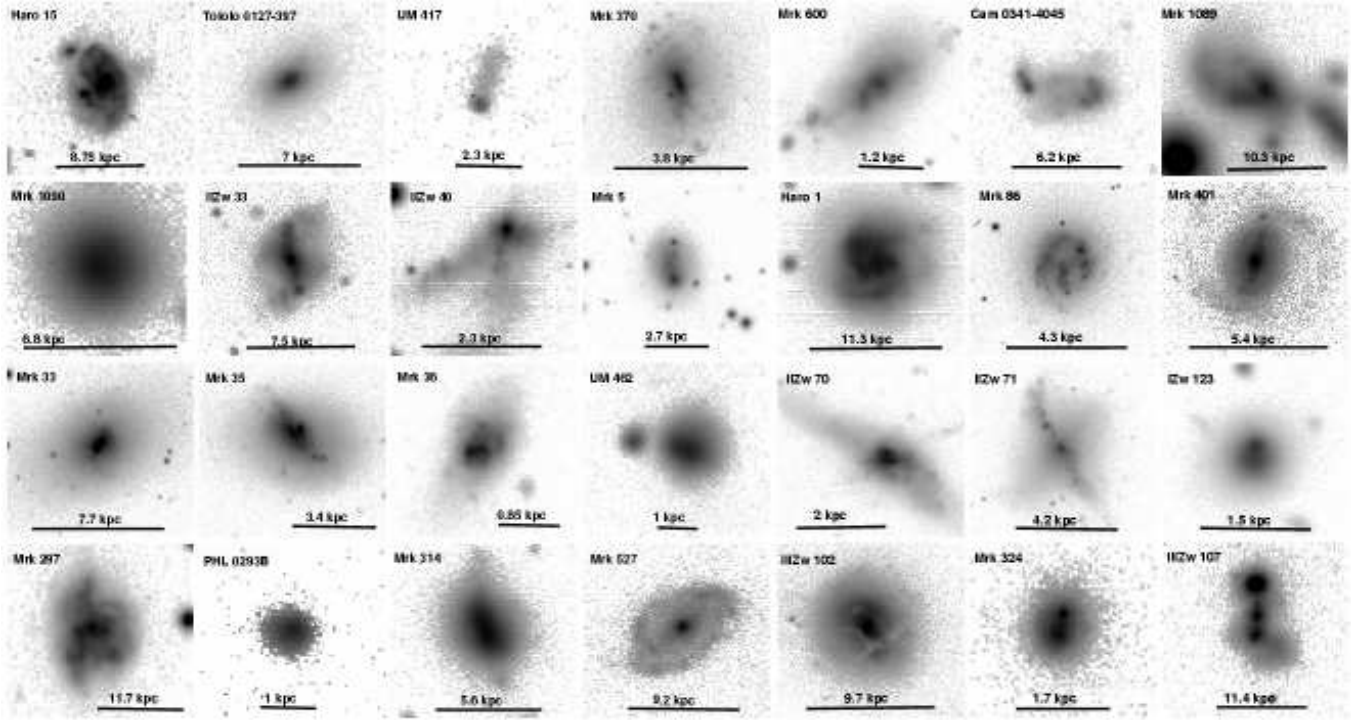}}
      \caption{The sample. $B$-band images are on a logarithmic greyscale. 
They are oriented in the standard N-E direction. 
              }
         \label{F1}
   \end{figure*}
%

$B$-band images of the galaxies are presented in Fig.~\ref{F1}.  
An inspection of the sample reveals that galaxies cover a wide range of luminosities 
($-21 \la$ M$_{\rm B}$ $\la -14$) and morphological BCG classes, such as 
those defined by \citet{LT86} from their optical morphologies. 
They go from small, compact objects dominated by a central star-forming region, to large and more 
extended objects in which star formation is spread out almost over the whole galaxy at high surface 
brightness levels. Moreover, several galaxies show particular morphological features such as bars, or 
typical signs of interactions, such as tails. Indeed, two galaxies (Mrk~527, Mrk~401) are 
luminous starburst galaxies rather than classical BCGs and show incipient spiral features (C01a). 

Table~\ref{T1} summarises some of the galaxies' basic data collected from previous studies. 
It includes distances and extinction coefficients that will be used in the present paper. 
Through this paper a Hubble constant of H$_0=$75 km s$^{-1}$ Mpc$^{-1}$ has been used.
%
%
\begin{table*}
\caption{Main parameters of the galaxy sample}
\label{T1}
\centering
\begin{tabular}{l c c c c c c c }
\noalign{\smallskip}
\hline\hline
\noalign{\smallskip}
Galaxy & RA(J2000) & Dec.(J2000) & D (Mpc) & $A_{\rm B}$ (mag)  &  $m_{\rm B}$ (mag) & $B-V$ & $M_{\rm B}$ (mag)  \\
(1)   &    (2)   &   (3)   &   (4)   &  (5)  &  (6)  &  (7)    &   (8)    \\
\noalign{\smallskip}
\hline
\noalign{\smallskip}
Cam~0341-404	& 03 42 49.4 &$-$40 35 56& 56.69& 0.068 & 16.62 & 0.36 & $-$17.21  \\[3pt]
Haro~1     	& 07 36 56.4 & 35 14 31	 & 52.11& 0.184	& 12.79 & 0.54 & $-$20.98  \\[3pt]
Haro~15    	& 00 48 35.9 &$-$12 43 07& 84.15& 0.101	& 14.06 & 0.35 & $-$20.67  \\[3pt]
Mrk~5           & 06 42 15.5 & 75 37 33  & 13.96& 0.364 & 15.58 & 0.60 & $-$15.51  \\[3pt]
Mrk~33     	& 10 32 31.9 & 54 24 03	 & 22.30& 0.052	& 13.46 & 0.41 & $-$18.33  \\[3pt]
Mrk~35          & 10 45 22.4 & 55 57 37  & 15.60& 0.031 & 13.21 & 0.61 & $-$17.79  \\[3pt]
Mrk~36          & 11 04 58.5 & 29 08 22  & 10.43& 0.131 & 15.38 & 0.35 & $-$14.84  \\[3pt]
Mrk~86          & 08 13 14.7 & 45 59 26  &  8.12& 0.232 & 12.41 & 0.73 & $-$17.37  \\[3pt]
Mrk~297    	& 16 05 12.9 & 20 32 32	 & 65.10& 0.330	& 13.16 & 0.44 & $-$21.24  \\[3pt]
Mrk~314         & 23 02 59.2 & 16 36 19  & 28.95& 0.383 & 14.15 & 0.07 & $-$18.53  \\[3pt]
Mrk~324 	& 23 26 32.8 & 18 16 00	 & 22.43& 0.215	& 15.39 & 0.58 & $-$16.58  \\[3pt]
Mrk~370         & 02 40 29.0 & 19 17 50  & 10.85& 0.399 & 13.59 & 0.53 & $-$16.99  \\[3pt]
Mrk~401	    	& 09 30 17.0 & 29 32 24	 & 24.07& 0.101	& 13.82 & 0.01 & $-$18.19  \\[3pt]
Mrk~527    	& 23 13 12.7 & 06 19 18	 & 47.56& 0.473	& 13.58 & 0.55 & $-$20.28  \\[3pt]  
Mrk~600     	& 02 51 04.6 & 04 27 14	 & 12.81& 0.282 & 15.29 & 0.31 & $-$15.53  \\[3pt]
Mrk~1089   	& 05 01 37.8 &$-$04 15 30& 52.65& 0.220	& 14.05 & 0.08 & $-$19.78  \\[3pt]
Mrk~1090   	& 05 01 44.1 &$-$04 17 19& 51.93& 0.238	& 15.16 & 0.10 & $-$18.65  \\[3pt]
PHL~0293~B    	& 22 30 33.9 &$-$00 07 35& 21.39& 0.302	& 17.65 & 0.51 & $-$14.30  \\[3pt]
Tololo~0127-397 & 01 29 15.8 &$-$39 30 37& 61.05& 0.067 & 16.18 & 0.51 & $-$17.81  \\[3pt]
UM~417      	& 02 19 30.2 &$-$00 59 11& 35.08& 0.141	& 18.04 & 0.29 & $-$14.83  \\[3pt]
UM~462     	& 11 52 37.3 &$-$02 28 10& 14.12& 0.083	& 14.69 & 0.47 & $-$16.14  \\[3pt]
I~Zw~123        & 15 37 04.2 & 55 15 48  & 12.52& 0.062 & 15.46 & 0.55 & $-$15.09  \\[3pt]
II~Zw~33   	& 05 10 48.1 &$-$02 40 54& 36.41& 0.439	& 15.02 & 0.81 & $-$18.22  \\[3pt]
II~Zw~40   	& 05 55 42.6 & 03 23 32	 &  9.69& 3.538	& 15.18 & 1.12 & $-$18.29  \\[3pt]
II~Zw~70        & 14 50 56.5 & 35 34 18	 & 19.12& 0.053 & 14.85 & 0.41 & $-$16.61  \\[3pt]
II~Zw~71   	& 14 51 14.4 & 35 32 32	 & 19.64& 0.055	& 14.46 & 0.56 & $-$17.06  \\[3pt]
III~Zw~102 	& 23 20 30.1 & 17 13 32	 & 22.71& 0.109	& 12.61 & 0.76 & $-$19.28  \\[3pt]
III~Zw~107	& 23 30 09.9 & 25 31 58  & 78.09& 0.259	& 15.08 & 0.60 & $-$19.64  \\[3pt]

\noalign{\smallskip}									      
\hline											      
\hline											      
\end{tabular}
\begin{list}{}{}
\footnotesize
\item Notes.$-$ Columns:  
(2) Right ascension in hours, minutes and seconds; 
(3) Declination in degrees, arcminutes and arcseconds; 
(4) Distance, computed assuming a Hubble flow with a Hubble constant 
$H_{0}$ = 75 km s$^{-1}$ Mpc$^{-1}$, after correcting recession velocities 
relative to the centroid of the local group for Virgocentric infall; 
(5) Extinction coefficient in the $B$ band, from \citet{schlegel98}; 
(6) Asymptotic magnitude in the $B$ band, from C01b. Note that the 
asymptotic magnitudes listed in C01b were corrected for Galactic extinction 
following \citet{BurstHeil84}; here they are presented uncorrected; 
(7) B$-$V colour from the asymptotic photometry of C01b, uncorrected for galactic extinction; 
(8) Absolute magnitude, obtained from the $B$ asymptotic magnitudes of (6) 
recomputed using the \citet{schlegel98} extinction values, 
distances were taken from those tabulated in (5); 
\end{list}
\end{table*}
%
\section{Fitting the BCG host galaxy: methodology}

To derive the structural parameters of the host, we adopted the two-dimensional fitting
 methodology developed and presented in detail in Paper~I. Our method uses the software 
{\sc galfit} v~2.0.3c \footnote{Detailed information on {\sc galfit} and its implementation 
can be found online, http://zwicky.as.arizona.edu/~cyp/work/galfit/galfit.html} \citep{Peng02} 
for fitting a two-dimensional model to the host galaxy light distribution. We refer 
 the reader to Paper~I for a detailed explanation of the fitting technique. A summary of the 
 main features of the method is given here, with some remarks about its implementation to this sample.

Based on the appearance of the surface brightness profile, C01a find for this sample that
approximately 50\% of the galaxies are objects with ``composite'' profiles that cannot be fitted by
a single exponential or $r^{1/4}$ law over the whole intensity range (see their Fig.~2). Most galaxies 
in the sample were described in C01a by an exponential model in their very outer profiles, 
i.e.\ the region where the light is dominated by the host galaxy. 
However, subsequent papers devoted to analysing the structure of BCGs by using this and other samples 
in the optical and NIR regimes, have shown that several BCGs may present 
light profiles with significant deviations from a pure exponential \citep[e.g.,][]{C03,Noeske03,Caon05}. 
In this project, and following our previous studies \citep[][and Paper~I]{Caon05}, we explored the 
suitability of the \mbox{S\'ersic} law \citep{Sersic68} to fit the host galaxy of our BCG sample. 
The \mbox{S\'ersic} law can be expressed as

\begin{equation}
I(r) = I_{e} e^{-b_{n}[(r/r_{e})^{1/n} - 1]},
\end{equation}
where $I_{\rm e}$ is the intensity at the effective radius \mbox{$r_{\rm e}$}, which 
encloses half of the total light from the model \citep{Caon93}. The constant $b_n$ is coupled 
with the \mbox{S\'ersic} index, $n$, and can be approximated as 2$n-1/3$ \citep[e.g.,][]{Graham01}. 
The S\'ersic law presents two particular cases, when $n=1$ and $n=4$; the first is a pure exponential 
model whereas the second is the $r^{1/4}$ law. The \mbox{S\'ersic} profile has been shown to describe 
the light distribution in ellipticals (from dwarfs to the brightest cluster members, 
\citep{Caon93,Graham96,G&G03,Gutierrez04,Aguerri04,Aguerri05a,Kormendy09}, LSB galaxies \citep{Cellone94}, 
and spiral bulges \citep{Andredakis,Prieto01,Aguerri03,Aguerri05b,Jairo07} 
besides BCGs \citep[e.g.,][; Paper~I; this paper]{C03,Caon05,Cairos07}. 
This way, \mbox{S\'ersic} models will allow us to explore a number of galaxy structures and 
compare their properties for the BCG hosts with those of other galaxy types. 

The 2D models fitted by {\sc galfit} v~2.0.3c are axially symmetric, generalised ellipses \citep{Lia90}, 
which have a centre ($x$,$y$) and a shape that is given by their axial ratio ($q=b/a$), position angle 
($PA$), and the boxy/discy parameter ($c$). This gives different shapes to the entire component depending 
on its values: $c<0$ (discy shape), $c = 0$ (pure ellipse), and $c>0$ (boxy shape).
In addition to the above five, the three \mbox{S\'ersic} free parameters, i.e.\ total apparent 
magnitude ($m_{\rm host}$), effective radius ($r_{\rm e}$), and \mbox{S\'ersic} index ($n$) were obtained 
by {\sc galfit} using a  nonlinear fitting procedure that minimises the $\chi^2_{\nu}$ distribution of 
the residuals (image $-$ model) weighting each pixel with its own noise, 
i.e.\ the Poisson error at the pixel position, after convolving models with a given PSF image. 
This procedure starts from an initial guess at the free structural parameters. 
 These values (except $c$, which was initially assumed as zero) were taken from previous fits of their 
1D surface brightness profiles \citep[see C01a; C01b;][]{Caon05}. 
To take the seeing of the images into account, the \mbox{S\'ersic} models were convolved by 
galfit by using a star selected from each image. 

Following the prescriptions given in Paper~I, a special effort was made to fit only 
those regions free of the starburst emission and all possible spurious sources 
that could contaminate the 2D models. 
Therefore, all background and foreground sources, cosmic rays and possible bad pixels
  were masked out from galaxy images by using the {\sc iraf/pros} task {\sl plcreate}.   
To mask out the starburst contribution from the fits, we made a first-guess set of elliptical
 masks by using the {\sc iraf/pros} tasks {\sl plcreate} and {\sl imreplace}. We obtained different 
 sized masks from the smaller ones following the inner isophotes, i.e.\ masking out the starburst
  peak(s), up to the larger ones, following the shape of the outer isophotes, i.e.\ as needed 
  to ensure that the entire starburst emission was completely masked out. Then, for a given 
  galaxy, the procedure generated several different models, one for each mask. In this way, we 
  analysed the influence of the starburst contamination on the \mbox{S\'ersic} parameters by 
  plotting the free parameters of each model versus the size of its mask, $R_{\rm mask}$.
From these plots we can usually distinguish $R_{\rm tran}$, i.e.\ the radius beyond which 
starburst emission is practically absent (see also P96a or C01b), and the range where the 
structural parameters are expected to be stable. 
Then, we used the positive residuals (galaxy$-$model) of those models where 
$R_{\rm mask} > R_{\rm tran}$ to refine the mask in shape and size in an iterative process. 
Finally, we obtained a new improved mask that follows the actual shape of the starbust 
emission. The final solution was obtained from those models fitted with the best improved mask, which 
gives the minimum $\chi^2_{\nu}$ residuals. Several examples that illustrate this procedure can be 
found in Paper~I.

\subsection{Error sources}
When fitting a \mbox{S\'ersic} law in BCGs, one may identify two main sources of uncertainty: 
the limited portion of galaxy able to be used in the fit and the sky subtraction uncertainties 
\citep[e.g.\ C03;][]{Noeske03,Noeske05}. In \citet{Caon05} and Paper~I, we analysed these 
drawbacks for both 1D and 2D BCG profiles, respectively. 
Although these error sources depend on the data quality, even when the same dataset was analysed, 
two-dimensional fits were found to be more robust for recovering the structural parameters 
when a limited number of pixels and surface brightness intervals were fitted. 
By using ideal simulations and the eight best quality and deepest galaxy images, we have 
shown that relative deviations of about 10--20\% are generally expected for low $n$ models 
when the radius of the masked region is between $\sim$1 and 2$r_{\rm e}$, whereas for higher 
$n$ values deviations increase considerably with the size of the mask. 

On the other hand, the fainter surface brightness levels of the galaxy images are affected by 
 inhomogeneities in the sky background. The noise and the sky background 
 subtraction uncertainties are propagated to the \mbox{S\'ersic} parameters, which could become 
 unstable, especially for higher $n$ values. This drawback, especially associated with the 
 \mbox{S\'ersic} profile, has been observed in several recent papers \citep[e.g.][Paper~I]{C03, Caon05}.
Two independent sky estimates were considered in our procedure. The first mean value was derived 
from several measurements of different boxes, free of sources, surrounding the galaxy (see C01a).
The second mean value was obtained by masking out all sources with a flux greater than $\sim$0.5 
times the mean {\em rms} of the background, and then fitting with {\sc galfit} only a polynomial 
function to the entire image (see Paper~I).
Generally, both results agree within $\la$ 0.5\% of the measured sky. In this way, the galaxies 
were fitted with a S\'ersic function plus a fixed sky function for both sky mean estimates, 
also considering the sky level as an additional free parameter. 
These three sets of free parameters were inspected to check the consistency among values 
and to estimate the uncertainty involved in the sky-subtraction.

Another systematic uncertainty affecting the fits comes from the galaxies 
not being perfect ellipsoids. This could be a remarkable weakness of the method, especially in 
BCGs with complex morphologies, where the derived structural parameters should be taken as 
their best average values. Our sample contains several BCGs showing isophotal twists and irregularities 
even in their outer regions (see e.g.\ II~Zw\,40), so they show large systematic residuals  
after fitting the 2D models. Most BCGs of the sample, however, belong to the morphological 
classes nE and iE proposed by \citet{LT86b} i.e.\ nuclear and irregular starburst, respectively, 
with an regular outer envelope, showing negligible systematic residuals in the fitted region. 

\subsubsection{Consistency checks}
To assess the reliability of the structural parameters, we need to carry out 
some consistency checks.
First, we analyse the accuracy on the $R_{\rm tran}$ estimation.
Several studies have revealed that this is crucial for avoiding overestimations in the \mbox{S\'ersic} 
models induced by the starburst contamination \citep[e.g.][]{Caon05,Noeske05}. 
Here $R_{\rm tran}$ is derived as the radius of a circle that contains the 
same number of pixels of the mask (see Paper~I). Initially, we compared the 
size of the mask with both the size of the $H_{\alpha}$ (net) emission
when narrow band data was available and with the size of the blue regions (starburst region) 
from the colour maps derived by C01b.
To avoid underestimations in $R_{\rm tran}$, we also compared its value with the radius 
at which the colour profile of the galaxy reached a nearly constant maximal value, characteristic of the 
old stellar component. 
We used colour profiles ($B-V$ and $V-R$) derived by C01a for the full sample of BCGs.
A set of $B-R$ colour profiles were also taken from GdP05 for a limited number of galaxies. 

Second, we explored whether a radius range exists where the fit is stable and 
does not depend on the exact choice of $R_{\rm tran}$. We show in Paper~I that galaxy profiles 
with $R_{\rm tran} \la 2 r_{\rm e}$ generally show a stability range.

Third, because we expect that the stellar host has negligible colour gradients 
\citep[as shown by the observed behaviour in the outer regions, see for example C01b;][]{Noeske05,GdP05}, 
both $n$ and $r_{\rm e}$ should be the same in all pass bands (within the errors), whereas the 
differences in $\mu_{\rm e}$ and $m_{\rm host}$ reflect the host galaxy colours. 
Finally, a reasonable agreement between ellipticity, position angle, and $c$ coefficient in 
different bands is desirable.  

\section{Results}

The results of the 2D fitting of a \mbox{S\'ersic} law to the starburst-free regions of the 
galaxy images, in the available filters, are presented in Table~\ref{T2} for the sample of 20 galaxies. 
Additionally, we included the results for the 8 galaxies fitted in Paper~I. 
Columns 3, 4, 5, 7, 8, and 9 show the free parameters fitted by {\sc galfit}: 
position angle ($PA$), axial ratio ($q$), boxy/discy parameter ($c$), the S\'ersic index ($n$), 
the effective radius ($r_{\rm e}$), and the total apparent magnitude of the model ($m_{\rm host}$), 
respectively. Column 6 shows the transition radius, $R_{\rm tran}$. Columns 10 to 13 show values 
inferred directly from the models: absolute magnitude of the model ($M_{\rm host}$), effective 
and central surface brightness ($\mu_{\rm e}$ and $\mu_{\rm 0}$) and the surface brightness at the 
transition radius ($\mu_{\rm tran}$). Finally, galaxies have been marked with a quality index, $Q$, as 
shown in Column 14. $Q=1$ indicates those galaxies (20/28) for which all consistency checks were 
fulfilled, while $Q=2$ indicates those galaxies (8/28) for which our 2D fitting by using a single 
S\'ersic model were not satisfactory.

\subsection{Reliability of the models}
Following Paper~I, in addition to the formal uncertainties derived by {\sc galfit} 
\citep[see][]{Peng02}, global uncertainties for $m_{\rm host}$, $r_{\rm e}$, and $n$ were 
computed by measuring their dispersion in the stability range, i.e.\ where $R > R_{\rm tran}$, 
and the difference between the values obtained when fitting a fixed sky-background and those obtained 
by setting the sky background level as a free parameter. These uncertainties are 
listed in Table~\ref{T2}. 
Likewise, the scatter of the shape parameters ($q$, $PA$ and $c$) and those of $n$ and 
$r_{\rm e}$ between the different filters could give us another qualitative measure of 
their uncertainty. 
For galaxies with $Q=1$, the scatter of the above parameters is as much as 20$-$25\%, always 
smaller than their global uncertainties, and can be explained in terms of the different quality.

To illustrate the 2D fits, broad-band images, \mbox{S\'ersic} models and the residual images 
for the full sample of 28 BCGs are presented in Figs.~\ref{Ap1} and 
\ref{Ap2}\footnote{Available in the online version}. For the sake of clarity we split the galaxies 
according to their quality index. A visual inspection of the residuals reveals those regions dominated 
by the starburst and those regions where the axially symmetric models cannot reproduce the 
host galaxy morphology, e.g.\ II~Zw\,71 or II~Zw\,40. 
In some galaxies it is also possible to distinguish complex features, such as spiral arms 
e.g.\ Mrk~401 and Mrk~527, dust lanes e.g.\ III~Zw\,102, or tails e.g.\ II~Zw\,70.  
Overall, morphologies of the host galaxies with $Q=1$ are mostly reproduced with the $q$, $PA$ 
and $c$ parameters. 

Twenty galaxies (8 from Paper~I, and 14 from this paper) show $Q=1$, fulfilling all the 
proposed consistency checks. For these galaxies we found $R_{\rm tran}$ values in good agreement
with the radii at which their colour profiles reach a nearly constant value. Small differences 
($\la$ 3 arcsec) may be attributed to the sometimes large error bars in the colour profiles and to the
different way to measure both radii. We also estimated the intervals of surface 
brightness and galaxy radius used in the fits as $\mu_{\rm out} - \mu_{\rm tran}$ and 
$R_{\rm out} - R_{\rm tran}$, where $\mu_{\rm out}$ and $R_{\rm out}$ are respectively 
the surface brightness and the radius of the outer reliable isophote, taken from the radial 
profiles presented in C01a. 
For $Q=1$ galaxies, a mean surface brightness interval of about 3.9 magnitudes and a 
coverage of the radial profile of about 67\% were obtained. 
Only three galaxies show $\mu_{\rm out} - \mu_{\rm tran} < 3$ magnitudes
(UM\,417, UM\,462, and Cam~0341-404), with a fitted radial range of about 50\%.
We notice, however, that these estimations are lower limits since those galaxy pixels
are fitted by {\sc galfit} down to the outer limits of the radial profiles presented in C01a. 
For $Q=1$ models the ratio $R_{\rm tran}/r_{\rm e}$ varies between 0.55 (UM~417\,$B$) 
and 1.85 (UM~462\,$V$).

Six of the eight galaxies with $Q=2$ did not pass the consistency checks for a number of reasons. 
The luminosity of the host model was greater than the luminosity of the whole galaxy 
(IIZw~33 and Mrk~1090). The fitted region was too small, i.e.\ $R_{\rm tran}$$\ga$2$r_{\rm e}$ 
(Haro~15, Haro~1, Mrk~297, IIIZw~107), and in the case of IIIZw~107; in addition, the light 
contamination caused by a bright foreground star located over the galaxy at the north of the image 
(see Fig.~\ref{F1}). In the remaining galaxies, two low surface brightness tails 
(IIZw~40) and the proximity of companions (Mrk~1089 and Mrk~1090) affect the stability 
of the fitted parameters, and no unique solution is obtained in these last cases 
(several solutions with similar $\chi^2$ value).

Seven of the eight galaxies with $Q=2$ have been reported in the literature as probable mergers or
 interacting galaxies: Mrk~297, a merging system 
 \citep[e.g.,][]{Smith96,Israel05}; Mrk~1089 and Mrk~1090, a merging system, which are members 
of the Hickson Compact Group 31 \citep{Hickson82,IglVil97,Rubin90}; Haro~15, strongly interacting 
separated galaxies or advanced merger \citep{Mazza91}; Haro~1, a paired galaxy \citep{Schnei92}; IIZw~40, 
 morphology interpreted as an interaction \citep{Bald82,BrinksKlein88} and merger of two gas-rich 
 dwarf galaxies \citep{Sage92,Joy&Lester,Deeg97}; and IIZw~33, morphology interpreted as an 
 interaction \citep{Mendez99}. Interestingly, six of the eight galaxies with $Q=2$ are 
 the most luminous ones in the full (28) sample. Excepts Mrk~1090, it is remarkable to note that
the above galaxies show the most irregular hosts in the sample. 

\subsection{Derived properties}  
%
\begin{table}
\begin{minipage}[t]{\columnwidth}
\caption{Mean structural parameters of the host for the sample.}
\label{T3}
\centering
\renewcommand{\footnoterule}{}  
\begin{tabular}{c c c c c}       
\hline\hline                 
 Parameter &   B   &  V   &  R   & Ref. \\

\hline 
$n$            &1.15 (0.55)&1.12 (0.44)&1.12 (0.45) & $a$  \\
$r_{\rm e}$    &1.11 (0.74)&1.05 (0.71)&1.09 (0.60)& $a$   \\
               &1.36 (1.23) & 1.22 (1.14) & 1.19 (0.72) & $b$ \\   
               &0.84& \nodata &1.06& $c$  \\
               &1.10& \nodata &1.25& $d$  \\
               & \nodata &0.83& \nodata & $e$  \\
$\mu_{\rm e}$  &22.59 (0.68)&22.19 (0.75)& 21.95 (0.76)& $a$   \\
               &23.56 (0.97) & 22.92 (0.98) & 22.53 (1.02) & $b$ \\
               &23.5&\nodata &22.8& $c$  \\
               &23.1&\nodata &22.5& $d$  \\
               & \nodata &22.8& \nodata & $e$  \\
\hline 
\hline                                  
\end{tabular}
\end{minipage}
\begin{list}{}{}
\item Notes. References: $a$$=$This work; $b$$=$C01a; $c$$=$\citet{GdP03}; $d$$=$\citet{Papade96a}; $e$$=$\citet{Hunter06}. Effective radii, $r_{\rm e}$, and surface brightnesses, $\mu_{\rm e}$, are expressed in kiloparsecs and magnitudes per square arcseconds.
Values in parentheses indicate the standard deviation for the sample.
  S\'ersic index for the $b$, $c$, $d$, and $e$ references may be considered as $n$$=$1, since 
in these papers a pure exponential profile was fitted to the host surface brightness distribution.
 \end{list}
\end{table}
%
We have derived several properties of the BCG sample concerning the two major components: the 
host and the starburst. In particular, we examined correlations among colours and structural 
parameters of the 2D host S\'ersic models (from Table~\ref{T2}) with quantities such as the H{\sc i} 
gas properties from the literature (see Table~\ref{T1}). Linear correlations were fitted taking 
the parameter uncertainties into account by using the routine {\sc fitexy} of \citet{Press92}. 
Those galaxies with ``poor'' fits ($Q=2$ in column 15 of Table~\ref{T2}) were not used in the 
following analysis. Galaxies in the well-fitted subsample of 20 are all ``classical'' dwarf 
systems according to their luminosities, i.e.\ $M_{\rm B}$$\ga$$-$19. 

Mean structural parameters and their standard deviations in {\em B, V}, and {\em R} are presented 
in Table~\ref{T3}. For comparison, some values from the literature were also
 added. Mean values for $\mu_{\rm e}$ and $r_{\rm e}$ agree, within the scatter, with 
 previous findings. However, slightly brighter mean $\mu_{\rm e}$ values have been obtained, probably the
 product of using a S\'ersic profile instead of pure exponential models (used by the other authors) 
 for the host component. 
The mean colours and standard deviations of the hosts are $B-V$$=$0.54$\pm$0.27 and $B-R$$=$0.95$\pm$0.26.

\subsubsection{Global colours}
The host colours have been derived from the 2D fits by using the extinction-corrected magnitudes 
of the S\'ersic model (see Table~\ref{T2}). Colours of the BCGs, i.e.\ derived from the 
asymptotic photometry (see C01b) and the colours of their hosts were compared. 
Overall, we found that the presence of a starburst component makes the BCG about 0.2 magnitudes 
bluer in $B-R$. Our BCG sample shows a wide range in the colour of their hosts ($\sim$1 mag). 
No clear correlation is  found, however, between the colours and structural parameters of the hosts. 

\begin{figure}
   \centering
   \includegraphics[width=8.5cm]{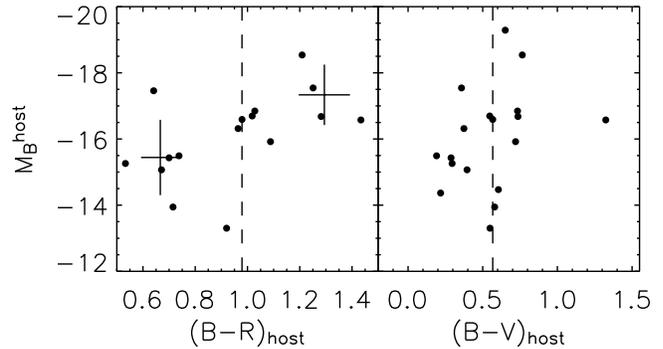}
   \caption{$B-V$ (right) and $B-R$ (left) colours versus the total $B$-band absolute magnitude of the host.
}
         \label{F2}
   \end{figure}

In Fig.~\ref{F2} we show the hosts distribution in the  $(B-V)-M_{B}^{\rm host}$ and 
$(B-R)-M_{B}^{\rm host}$ diagrams. The colours do not correlate with luminosity. However, three groups 
of galaxies may be roughly differentiated by their colours, especially in $B-R$. While several 
galaxies lie near the median (dashed line), in the extremes of the colour distribution we see both a 
relatively ``red'' group of galaxies, namely Tololo~0127-397,Mrk~86, Mrk~324, Mrk~33, II~Zw~123, and IIIZw~102, 
and a ``blue'' group, namely UM~462, UM~417, Mrk~600, Mrk~5, Mrk~401, PHL~0293B, Cam~0341-404, and II~Zw~70. 
The mean colours and standard deviations for the ``red'' and ``blue'' groups are $<$$B-V$$> =$ 0.86$\pm$0.26 
and 0.30$\pm$0.08, and $<$$B-R$$> =$ 1.29$\pm$0.10 and 0.66$\pm$0.07, respectively. In Fig~\ref{F2} 
(left) we added two crosses indicating these mean values for the two groups.

%
%
%
%
   \begin{figure}
   \centering
   \includegraphics[width=8cm]{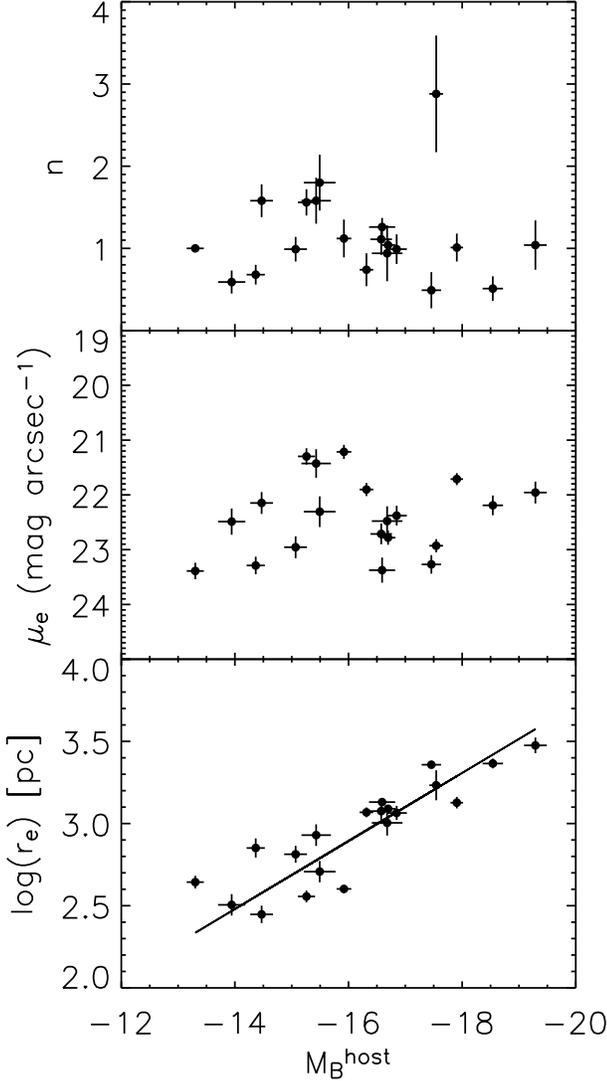}
      \caption{The \mbox{S\'ersic} index ({\em top}), effective surface brightness ({\em centre}), 
      and the effective radius ({\em bottom}), versus the $B$-band absolute magnitude of the host.
              }
   \label{F4}
   \end{figure}
%
\subsubsection{Correlations among the structural parameters}
To explore trends among the host structural parameters and their luminosities, 
Fig.~\ref{F4} shows the S\'ersic parameters $n$, $r_{\rm e}$, and $\mu_{\rm e}$ versus the $B-$band 
absolute magnitude (from Table~\ref{T2}). 
In the upper plot of Fig.~\ref{F4} we see that all the galaxies but one (Mrk~33) show hosts with 
\mbox{S\'ersic} indexes $n$$\la$2. Although the scatter is large, an approximately constant mean value 
$n$$\sim$1 is obtained in the whole range of host luminosities (see Table~\ref{T3}). We see that 
$\sim$50\% of the galaxies are well-fitted by ``pure'' exponential light distributions ($n$$\sim$1). 
The other 50\%, however, show more than $\sim$30\% deviations from the pure exponential model. 

No clear correlation between $\mu_{\rm e}$ and $M_{\rm host}$ was found, as shown in Fig.~\ref{F4} 
(centre). The sample spans  more than two magnitudes in effective surface brightness over the whole range 
of luminosity, with a constant behaviour similar to $n$. The same result, although with wide scatter, is 
obtained when the extrapolated central surface brightness is plotted. 

In contrast, $r_{\rm e}$ and $M_{\rm host}$ are closely correlated, as shown in Fig.~\ref{F4} (bottom). 
Linear fits to the data give
\begin{equation}
\log(r_{\rm e})[pc] = -0.21 \ (\pm 0.01) \ M_{\rm host} - 0.42 \ (\pm 0.14)
\end{equation}
with a Spearman rank index of $\rho = -$ 0.88. From this, we may conclude that the more luminous hosts are more
 extended. Similar trends were found for dwarf galaxies, including BCGs, by other authors in the optical 
 \citep[e.g.,][]{Vennik,Bergvall99} and in the NIR regime \citep[e.g., P96b;][]{Papade02,Vadu05,Vadu06}. 

The above results will be discussed in \S5 when a comparison is made with other galaxy types.
 
\subsubsection{Luminosities and sizes of the starburst and it host galaxy}
Assuming that the total luminosity of a BCG comes exclusively from two galaxy components, i.e.\ the host 
(described by the S\'ersic law) and the starburst episode, we computed the starburst luminosity as 
$L_{\rm SB} = L_{B}^{\rm tot}-L_{B}^{\rm host}$, where $L_{\rm tot}$ was taken from C01b (see Table~\ref{T1}). 
In addition, we adopted the transition radius as an estimate of the starburst average size.

\begin{figure*}
   \begin{tabular}{c c c}
   \includegraphics[angle=0,width=5.5cm]{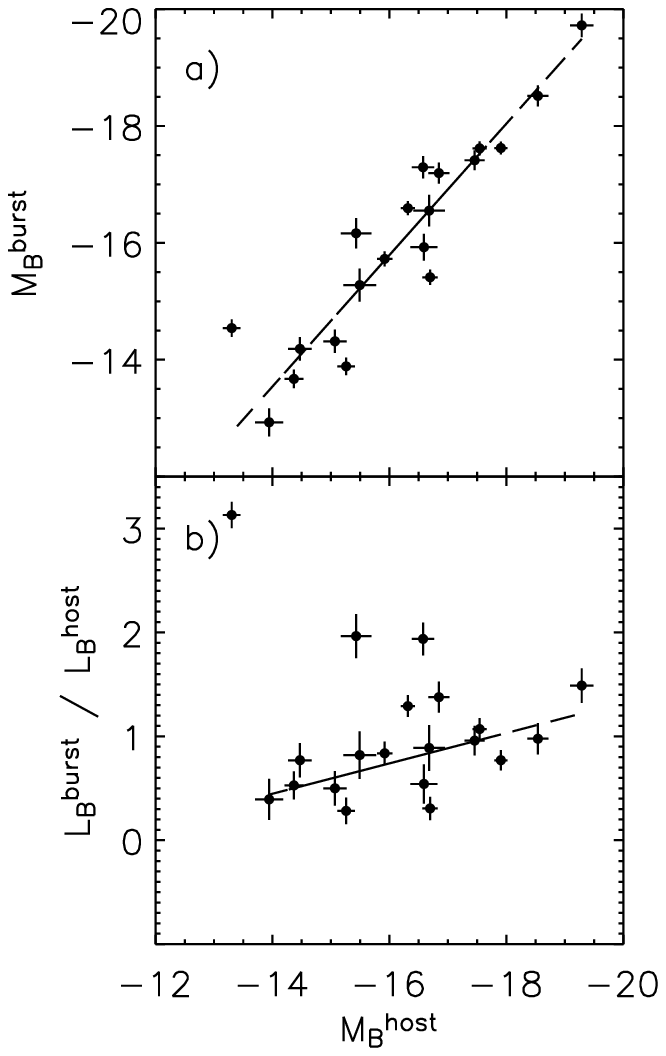} &
   \includegraphics[angle=0,width=5.35cm]{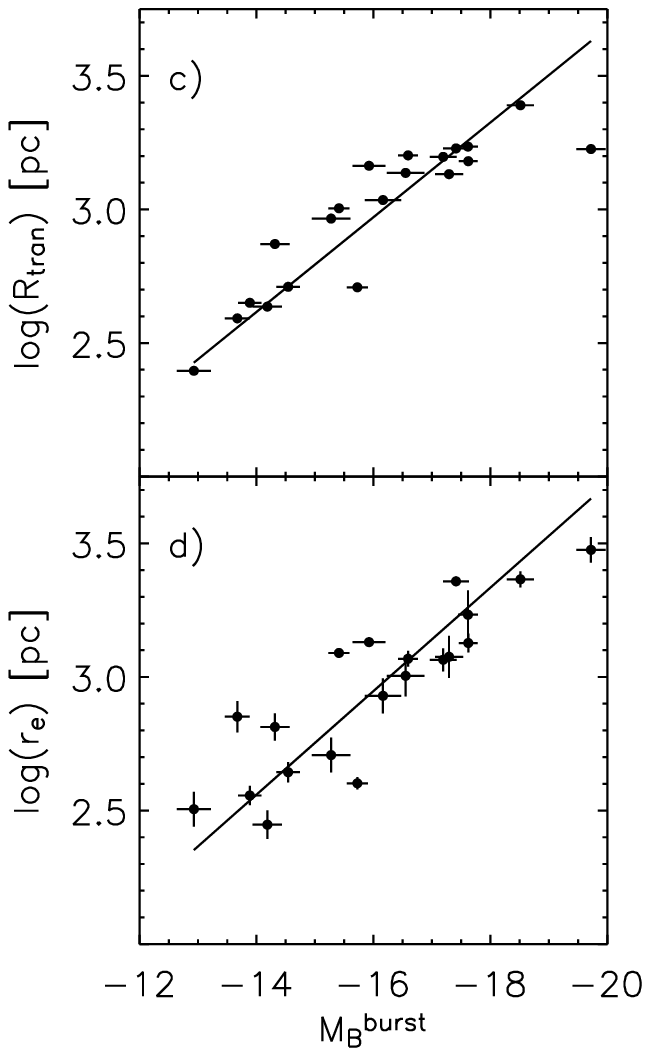} & 
\includegraphics[angle=0,width=5.20cm]{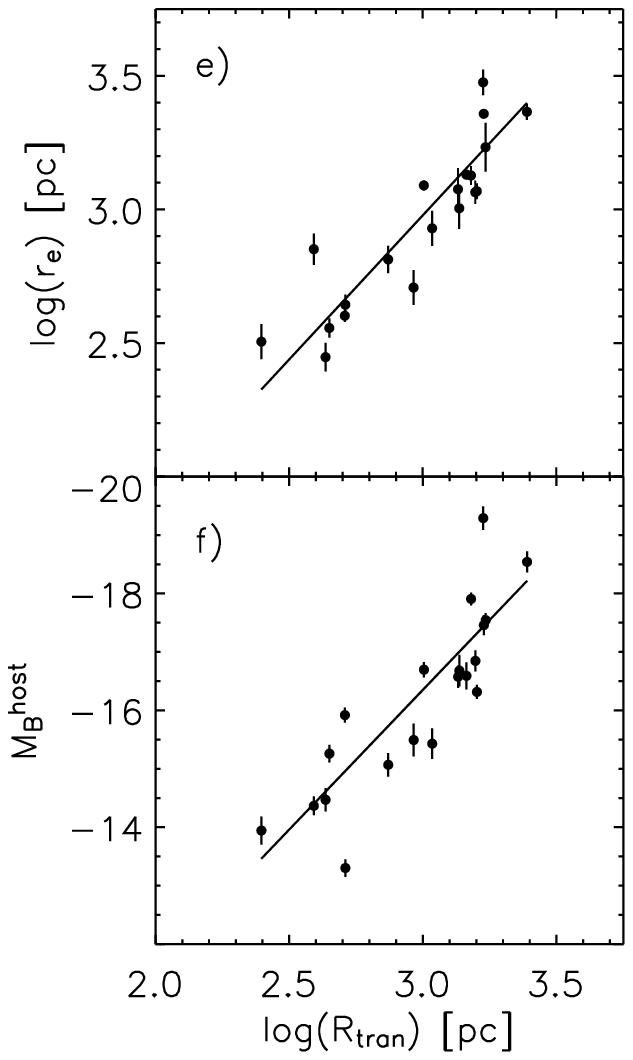}   
\end{tabular} 
   \caption{Sizes and luminosities of the two BCG components: a) $B$-band absolute magnitude of the starburst;
 and b) ratio between the starburst luminosity and the luminosity of the host as a function of the $B$-band absolute 
magnitude of the host; 
c) and d) transition radius and effective radius of the host versus the $B$-band absolute magnitude of the starburst;
e) and f) $B$-band effective radius and absolute magnitude of the host versus the transition radius. The solid 
lines show linear correlations to data. In b), the weighted fit includes only galaxies with $L_{\rm SB}/L_{\rm host}<$ 1.8.}
   \label{F5}
   \end{figure*}

Figure~\ref{F5}$a$ shows the strong correlation between the starburst and host $B$-band absolute magnitudes.
 We see that galaxies with more luminous hosts show luminous starbursts. Their relative strength, i.e.\ the
  ratio between the luminosities of both hosts and starbursts, is plotted versus the host blue absolute magnitude 
  in Fig.~\ref{F5}$b$. There is  large scatter, and a very weak correlation ($\rho=-0.36$) is noticed only when 
  the outlier (Mrk~36) is excluded. Thus, there is no clear evidence that the relative strength of the starburst 
  scales with galaxy mass. This result agrees with \citet{Vadu06} (see their Fig.~8), who did not see correlation
   between the relative strength of the starburst and the $K$-band luminosity of the host. 
Here, the mean ratio $<$$L_{\rm SB}/L_{\rm host}$$> =$ 1.15 $\pm$ 0.65 
   (0.93$\pm$0.5 when excluding the outlier) is equivalent to saying that, when the host galaxy suffers a 
   starburst episode, its $B$-band luminosity increases by $\sim0.77$mag. In the optical, we found a 
   wide range of values for $L_{\rm SB}/L_{\rm host}$ in the literature \citep[e.g.][; P96b]{Drink91,Meurer92,Taylor94}. 
   Our values are in good agreement with those of P96b.

In Figs.~\ref{F5}$c$ and \ref{F5}$d$ we have plotted the effective radius and the transition radius 
versus the $B$-band absolute magnitude of the starburst, while Figs.~\ref{F5}$e$ and \ref{F5}$f$  show
the host effective radius and $B$-band absolute magnitude versus the transition radius. 
All previous quantities are correlated,  their linear fits being given by
\begin{equation}
\log(r_{\rm e}) = 1.08 \ (\pm 0.03) \ \log(R_{\rm tran}) - 0.26 \ (\pm 0.10)
\end{equation}
\begin{equation}
\log(R_{\rm tran})[{\rm pc}] = -0.18 \ (\pm 0.01) \ {M_{\rm B}}^{\rm burst} - 0.14 \ (\pm 0.09)
\end{equation}
\begin{equation}
{M_{\rm B}}^{\rm host} = -4.78 \ (\pm 0.14) \ \log(R_{\rm tran})[{\rm pc}] - 2.00 \ (\pm 0.43)
\end{equation}
\begin{equation} 
\log(r_{\rm e})[{\rm pc}] = -0.19 \ (\pm 0.01) \ {M_{\rm B}}^{\rm burst} - 0.15 \ (\pm 0.13) 
\end{equation}
with Spearman's rank indexes $\rho$$=$0.89, $-$0.93, $-$0.88, and $-$0.86, respectively. 
The strong correlations tell us that the average size and the luminosity of the starburst emission does 
depend on the extent of the galaxy. Furthermore, since more luminous galaxies show more luminous starbursts, 
clear linear relationships between both $r_{\rm e}$ and $R_{\rm tran}$ with the starburst luminosity are 
found. These results support the picture presented by P96b: although star formation in BCGs should occur 
at random locations, the total average size of the star-forming region does depend on the extent of the host. 
This suggests that, therefore, the starburst episode in this sample of BCGs is regulated
 by the galaxies themselves on large scales, and does not depend on external issues.  
%
\subsubsection{Structure and gas content}
For those galaxies for which H{\sc i} mass data were available in the literature,
we explored possible correlations between the gas content of the galaxies and their optical sizes and luminosities.
Data on H{\sc i} gas mass and line widths, as well as the H{\sc i} gas mass-to-$B$--band host luminosity ratio, are 
listed in Table~\ref{T4}. For the sake of consistency, H{\sc i} masses from the literature were re-calculated by 
using the same distances as for estimating the optical absolute magnitudes (see Table~\ref{T1}). 

\begin{figure*}
    \begin{tabular}{l l l}
\includegraphics[angle=0,width=5.35cm]{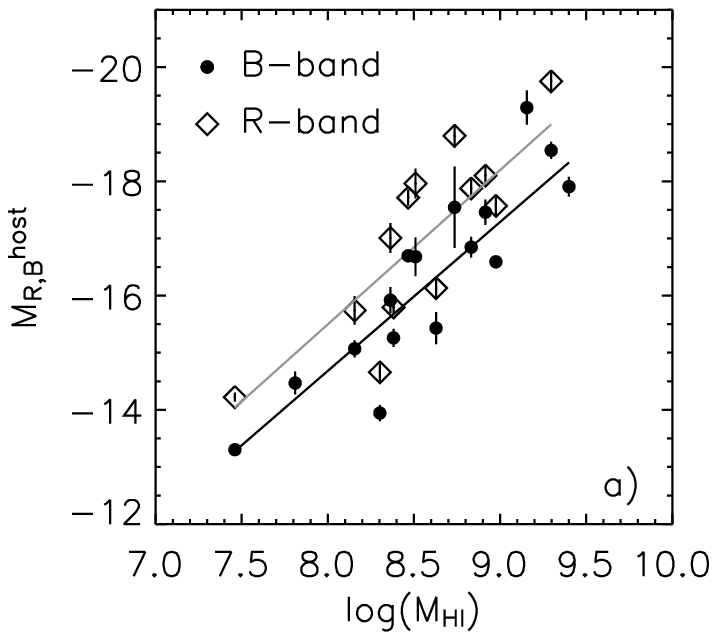}&
\includegraphics[angle=0,width=5.50cm]{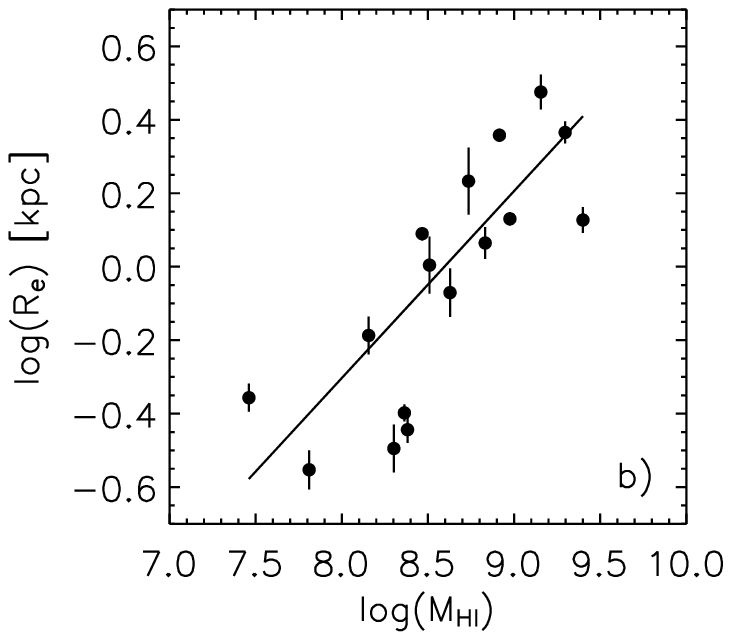}&
\includegraphics[angle=0,width=5.50cm]{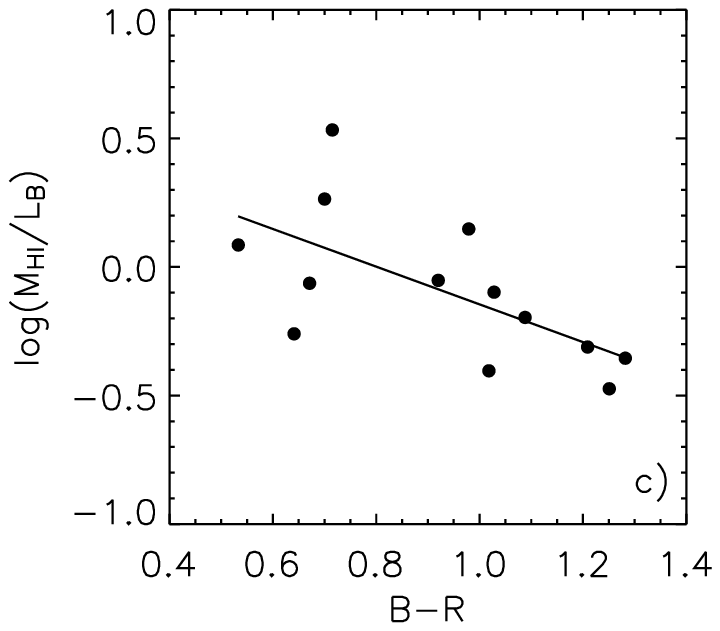}
    \end{tabular}   
    \caption{Structure of the host and the gas content of the BCG sample: a) $B$ (grey line) 
    and $R$ (black line) absolute magnitudes of the host versus the log of the neutral hydrogen 
    mass; b) $B$-band effective radius versus the neutral hydrogen mass; c) the ratio between the 
    H{\sc i} mass and $B$-band luminosity of the host versus the $B-R$ colour of the host. 
    }
   \label{F6}
   \end{figure*}

\begin{table*}[t]
\caption[]{HI gas and stellar properties for the analysed galaxies}
\label{T4}
\centering
\begin{tabular}{ l c c c c c }
\noalign{\smallskip}
\hline\hline
\noalign{\smallskip}
Galaxy & $W^{\rm c}_{\rm 20}$ &      M$_{\rm HI}$    &      M$_{s}/L_{B}$     &       M$_{\rm s}$   & M$_{\rm HI}/L_{\rm B}$ \\[3pt]
        &   [km s$^{-1}$]       & [10$^9$~M$_{\odot}$] & [M$_{\odot}$$/L_{\odot}$] & [10$^9$~M$_{\odot}$]&  [M$_{\odot}$$/L_{\odot}$] \\
    (1) &  (2)  & (3)  & (4)  &  (5)  &  (6)  \\
\noalign{\smallskip}
\hline
\noalign{\smallskip}
Mrk~5           &   90 & 0.14 & 0.41  & 0.07 & 0.86 \\[3pt]
Mrk~33     	&  196 & 0.54 & 2.19  & 3.55 & 0.34 \\[3pt]
Mrk~35          &  135 & 0.68 & 1.15  & 0.98 & 0.80 \\[3pt]
Mrk~36          &   78 & 0.03 & 0.85  & 0.03 & 0.87 \\[3pt]
Mrk~86          &   87 & 0.32 & 2.40  & 1.75 & 0.44 \\[3pt]
Mrk~314         &  269 & 2.51 & 0.09  & 0.21 & 1.11 \\[3pt]
Mrk~324 	&  135 & 0.23 & 1.37  & 0.50 & 0.67 \\[3pt]
Mrk~370         &  187 & 0.29 & 1.12  & 0.83 & 0.40 \\[3pt]
Mrk~401	    	&  148 & 0.82 & 0.38  & 0.57 & 0.55 \\[3pt]
Mrk~600     	&  102 & 0.24 & 0.28  & 0.05 & 1.22 \\[3pt]
PHL~293~B    	&  124 & 0.20 & 0.47  & 0.03 & 3.41 \\[3pt]
I~Zw~123        &  105 & 0.06 & 1.25  & 0.12 & 0.68 \\[3pt]
II~Zw~70        &  123 & 0.43 & 0.45  & 0.10 & 1.84 \\[3pt]
II~Zw~71   	&  180 & 0.95 & 1.00  & 0.65 & 1.41 \\[3pt]
III~Zw~102 	&  183 & 1.98 & 1.94  & 7.88 & 0.49 \\[3pt]
\noalign{\smallskip}									      
\hline											      
\hline											      
\end{tabular}
\begin{list}{}{}
\item Notes.$-$ Columns:
(2) Hydrogen line width at 20\% after inclination and random motion corrections (see \S5.2).(3) HI mass after correcting from differences in galaxy distances (see text). The original quantities of (2) and (3) were taken from \citet{ThuanMartin81} excepting Mrk~36, Mrk~324, Mrk~370 and III\,Zw~102 for which line width data was taken from \citet{Spring05}. 
(4) Mass-luminosity relation derived for the BCG hosts (see \S5.2); 
(5) Stellar mass, estimated from the host absolute magnitudes (see Table~\ref{T2}) and the $M/L$ values listed in (4); 
(6) HI gas mass-to-$B$--band host luminosity ratio. 
\end{list}
\end{table*}
%
In Fig.~\ref{F6}$a$ we plot the $B$ and $R$-band absolute magnitudes of the host versus the 
neutral hydrogen mass, $M_{\rm HI}$. Linear correlations in both filters give
\begin{equation}
M_{\rm B}^{\rm host} = -2.61 \ (\pm 0.04) \ \log(M_{\rm HI})[M_{\odot}] + 6.16 \ (\pm 0.36)
\end{equation}
\begin{equation}
M_{\rm R}^{\rm host} = -2.70 \ (\pm 0.07) \ \log(M_{\rm HI})[M_{\odot}] + 6.09 \ (\pm 0.60)
\end{equation}
($\rho=-$0.89 and $-$0.84, respectively). These correlations follow the same trend as those found by
 \citet{Stave92} (see their Fig.2) for the $B$-band luminosity of the entire galaxy (host$+$starburst). 
Moreover, since sizes and luminosities are correlated 
\citep[see Fig.~\ref{F4} and ~\ref{F5}, see also][]{Stave92,Campos93} a relationship between sizes (of both the 
starburst and the host) and gas content should be expected. By  simultaneously fitting the starburst and the
 host components of 12 BCGs, P96b found a correlation between the starburst diameter and the total H{\sc i} 
 mass (see their Fig.~3). A similar correlation is observed here between $M_{\rm HI}$ and $r_{\rm e}$, as
  shown in Fig.~\ref{F6}$b$. The linear fit gives
\begin{equation}
\log(R_{\rm e}) [{\rm kpc}] = 0.51 \ (\pm 0.02) \log(M_{\rm HI}) [M_{\odot}] - 4.37 \ (\pm 0.15) 
\end{equation}
($\rho=$ 0.85). The above correlation confirms the result obtained by P96b, showing that galaxies with
 larger gas reservoirs are more extended, and they also show extended starbursts (see Eq.\ 4). 

The H{\sc i} mass-to-blue luminosity ratio, $M_{\rm HI}/L_{\rm B}$, does not depend on the 
size of the galaxy components but on its colours. 
In Fig.~\ref{F6}$c$ we plot the ratio between the H{\sc i} mass and the 
$B$-band luminosity versus the $B-R$ colour of the host. The linear fit to the data gives
\begin{equation}
\log(M_{\rm HI}/L_{\rm B}) = -0.81 \ (\pm 0.24) \ B-R + 0.62 \ (\pm 0.24)
\end{equation}
($\rho=$ 0.63). Despite the large scatter, Eq.~11 suggest that those BCGs with redder hosts 
show smaller gas fractions than those with bluer ones. 
%
%
%
\section{Discussion}

\subsection{The host galaxy of BCGs and  possible links with other galaxy types}

Several common evolutionary scenarios involving dwarf galaxies types, i.e.\ BCGs, dIs and dEs, 
have been proposed and discussed in the literature in the last thirty years 
\citep[see e.g.,][]{Kunth&Ostlin}. Some authors \citep[e.g.][]{Thuan85,DyPh88} have suggested 
that BCGs could be the link connecting the possible evolution between  gas-rich dIs and  
gas-poor dEs, i.e.\ dIs should evolve into dEs after repetitive BCG stages. Similarities between 
dIs and BCGs that support this scenario have been found from their NIR colours \citep{Thuan85}, 
H{\sc i} properties \citep{Stave92} and spatial distributions \citep{Pusti95}. Moreover, some 
authors \citep{C00,ByO02} have found that the structural parameters of  BCG hosts are similar 
to those of other dwarfs. For instance, while BCGs and dIs in the Virgo cluster seem to be similar 
structurally and dynamically \citep{Vadu06}, at least $\sim$15\% of the BCGs in the large sample
 of \citet{GdP05}  could be dEs experiencing a burst of star formation. They also 
 found that BCGs with redder hosts ($B-R \ga$1 mag) show structural properties compatible with
  those of dEs.  

In contrast, systematic differences between their optical structural parameters led P96b to 
argue that evolutionary connections between BCGs, dEs, and dIs are  possible only if the BCG hosts  
  are able to change their  optical structural properties substantially. A more recent 
extensive comparison between the structural properties of BCGs and dIs by \citet{Hunter06} agrees 
with this claim. Furthermore, the observed neutral gas dynamics put strong constraints on the
 evolutionary fate of BCGs. Based on gas kinematics, \citet{Vanzee01b} conclude that BCGs 
 cannot evolve passively into dEs.  

Several of the above results were often obtained with a limited number of objects, different 
selection criteria, diverse analysis techniques (e.g.\ for subtracting the starburst emission), 
or with data of insufficient quality. Thus, no definitive conclusions about a unified evolutionary scheme 
for dwarf galaxies has been settled.  

In this paper we analysed the structural properties of a relatively large sample of
 BCG hosts by using a 2D fitting technique on deep optical images. In particular, we 
 investigated how the host in our BCG sample relates with samples of different types of galaxies
  in the diagrams built upon the \mbox{S\'ersic} parameters. 

\begin{figure} 
\centering
   \includegraphics[angle=0,width=8cm]{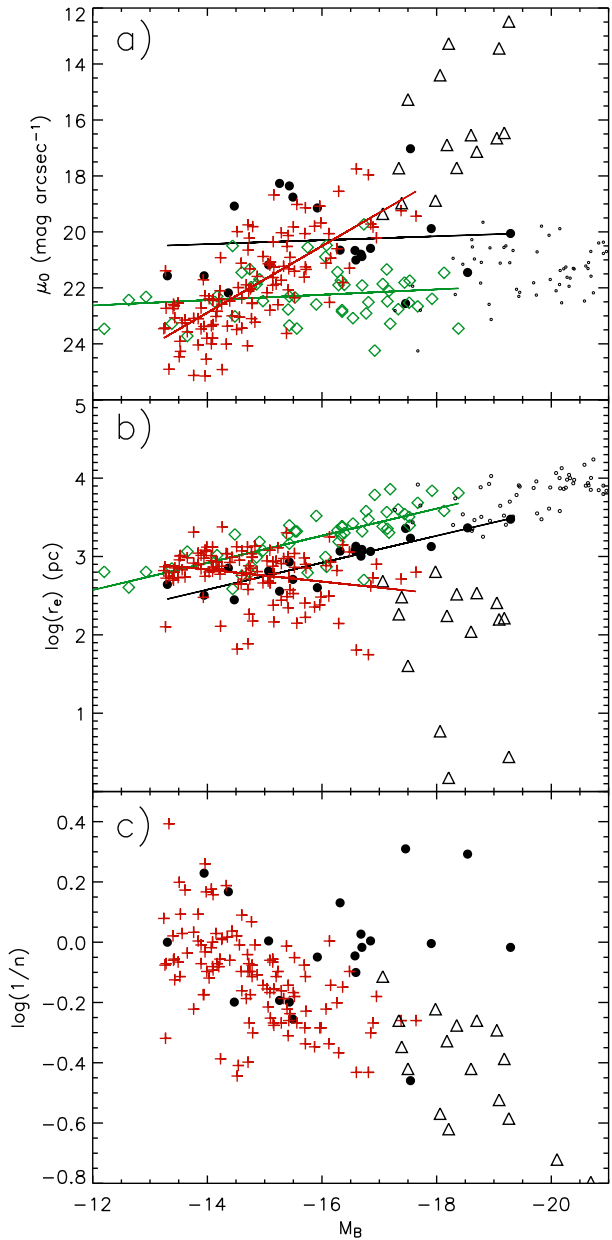}
      \caption{$(a)$ Central surface brightness, $(b)$ effective radius, and $(c)$ 
      \mbox{S\'ersic} index versus the $B-$band absolute magnitude of different galaxy types.
       Black circles and black lines represent the BCG host data and linear fits of our sample. 
       Green diamonds and lines, as well as red crosses and lines, show the dIs and dEs data and
        linear fits, respectively. Triangles show E-S0 data, whereas small dots indicate spiral-disc data.        
      }
   \label{F7}
   \end{figure}

Figures~\ref{F7}$a$,~\ref{F7}$b$, and~\ref{F7}$c$ show the $B$-band central surface brightness 
($\mu_{\rm 0}$), the effective radius ($r_{\rm e}$), and the \mbox{S\'ersic} index ($n$) of the 
host as a function of  absolute $B$-band magnitude, respectively. We also plotted 
 data for other galaxy types: E/S0 galaxies \citep[from][]{Caon93,donofrio94}, dEs in
 Virgo \citep[from][]{Bingelli98} and in Coma 
 \citep[from][]{G&G03}, isolated dIs \citep[from][]{Vanzee00}, and the exponential disc models of 
 \citet{Graham03} for the 36 spirals of \citet{Jong&Kruit94}. 
As shown in \S4.1, the central surface brightnesses and S\'ersic $n$ parameters are nearly constant 
over the full range of luminosities. In contrast,  effective radius is strongly correlated 
with the $B$-band absolute magnitude. Over the full range of host luminosities, i.e.\ 
$-13 \ga M_{\rm B} \ga -19$ these trends are different from those observed for dEs and Es, but 
similar to the trends shown by dIs and spiral discs.  
However, when inspecting the mean values we found clear differences, especially between BCGs 
and dIs. The mean $B$-band central surface brightness and the mean effective radius are 
20.28$\pm$1.44 mag and 1.11$\pm$0.74 kpc, respectively, two magnitudes brighter than and twice as 
large as those for Van Zee's isolated dIs ($<$$\mu_{\rm 0,B}$$> =$ 22.27$\pm$0.92 mag and
 $<$$r_{\rm e}$$> =$ 2.25$\pm$1.72 kpc).
%
%
%
Despite the very different fitting techniques used for deriving the host structural 
parameters, the above result for BCGs and dIs are in good agreement with previous 
findings \citep[e.g., P96b;][]{Hunter06}. Therefore, our results support the picture 
presented by P96b: if dIs evolve into dEs through repeatitive BCG phases, the hosts must 
undergo significant changes in their structure. 

On the other hand, we show in \S4.1.1 that, based on the global colours, two groups may 
be distinguished in our sample from those galaxies with colours close to the median. 
Comparing their mean colours with those of dIs and dEs, we found that the ``reddest'' hosts 
show mean global colours $<$$B-V$$> =$ 0.86$\pm$0.26 and $<$$B-R$$> =$ 1.29$\pm$0.10 
compatible with the red hosts of \citet{GdP05} and very similar to typical colours for dEs. 
For instance, \citet{Vanzee04a} derive $<$$B-V$$>= $0.77 and $<$$B-R$$> =$1.26 for a sample of dEs in Virgo. 
In contrast, the ``blue group'' of BCG hosts show mean global colours 
$<$$B-V$$> =$ 0.30$\pm$0.08 and $<$$B-R$$> =$ 0.66$\pm$0.09, very similar to the 
typical mean colours of  dI galaxies, e.g.\ $<$$B-V$$> =$ 0.42$\pm$0.05 for a large 
sample of isolated dIs \citep{Vanzee00}.

Although BCG hosts show clear differences in their average structural properties from 
those of dIs and dEs, a more individualised inspection of their properties shows that 
in the wide range of parameter values, morphologies, and colours, this kind of analysis 
cannot exclude the possibility that some galaxies in the sample (and in the general BCG 
``zoo'') could be  connected in an evolutionary sense with the dI or dE families. In particular, the 
reddest ones, which generally show smooth morphologies and central star formation 
(nE BCGs in the usual nomenclature), could be associated with some family of dE galaxies 
 \citep[see also][]{GdP05} sharing some of their properties. Nucleated dEs or dE 
 with blue centres, recently found in an almost-complete sample in the Virgo cluster by 
 \citet{Lisker06}, could probably be a good template for future studies linking dEs and BCG hosts 
 with regular morphologies and red colours.

\subsection{The Tully--Fisher relation}

%
   \begin{figure}
     \centering
       \includegraphics[angle=90,width=7.6cm]{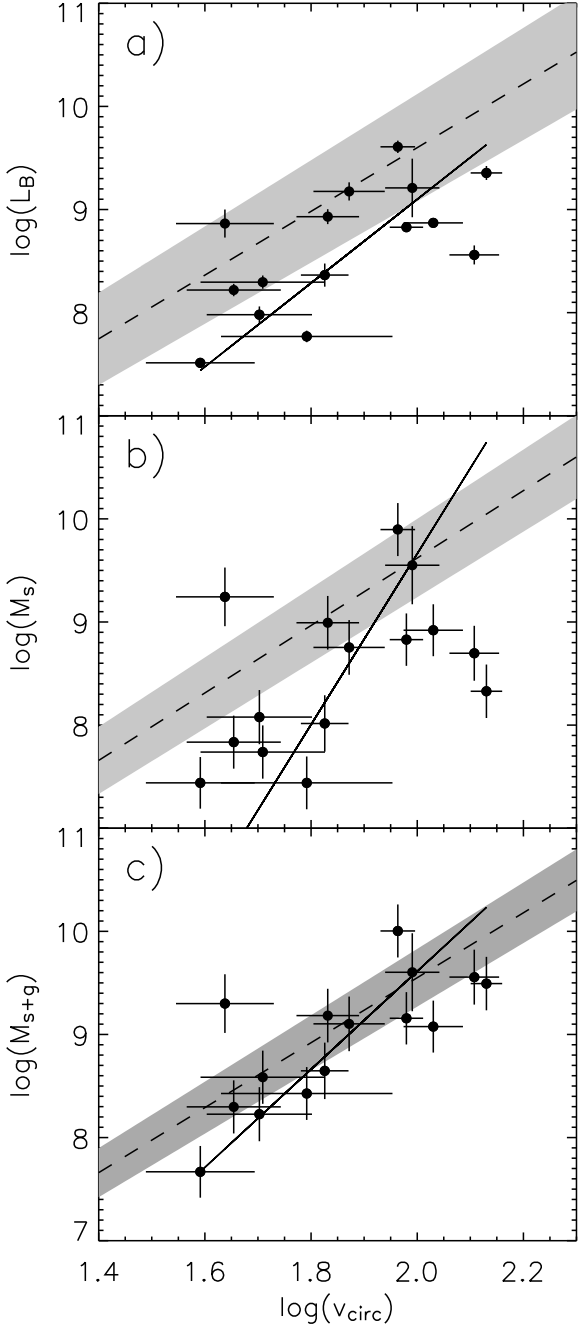}
     \caption{Tully--Fisher relations for the host (dots). $(a)$ TFR, the $B$-band luminosity 
     is plotted against the circular velocity. $(b)$ sTFR, the stellar mass is plotted 
     against the circular velocity. $(c)$ gsTFR, the H{\sc i} gas$+$stellar mass versus 
     the circular velocity. Solid lines indicate linear fits of the TFR, sTFR, and gsTFR 
     for our BCG hosts. Dashed lines indicate linear fits of the TFR, sTFR, and gsTFR for
      late-type disc galaxies  \citep[from][]{DeRick07}. The grey area indicates the 1$\sigma$ 
      level of the relations derived by \citet{DeRick07}. 
     }
     \label{F8}
   \end{figure}
%
The intrinsic luminosity $L$ and the maximum rotational velocity $V_{\rm max}$ are
 empirically related by the so-called Tully--Fisher relation \citep[TFR,][]{TF77}. This
  relation has been derived for bright spiral galaxies, for which it is assumed that the
   relation is linear in a logarithmic sense. However, the TFR had also been found to hold 
   for several galaxy types, including low surface-brightness spiral galaxies \citep{Zwaan95},
    dwarf late-type galaxies, i.e.\ Im-BCD galaxies 
    \citep[e.g.,][]{StilIsrael98,PierTuffs99,Pier99,Vadu05}, and early-type galaxies \citep{DeRick07}. 
    Moreover, for gas-rich, late-type galaxies, the TFR has been generalised to a
     relation between $V_{\rm max}$ and the stellar mass, M$_{\rm s}$ (sTFR), as 
     well as $V_{\rm max}$ and the H{\sc i} gas$+$stellar mass, M$_{\rm s+g}$, usually 
     called the ``baryonic Tully-Fisher relation'' (gsTFR) \citep{Mcgaugh00}, which could serve 
     for analysing possible environmental effects such as loss of gas mass in dEs \citep{DeRick07}. 
     We analysed the position of the BCG hosts in these scaling relations, 
     comparing them to other galaxy types.

%
   \begin{figure}
     \centering
       \includegraphics[angle=90,width=8.cm]{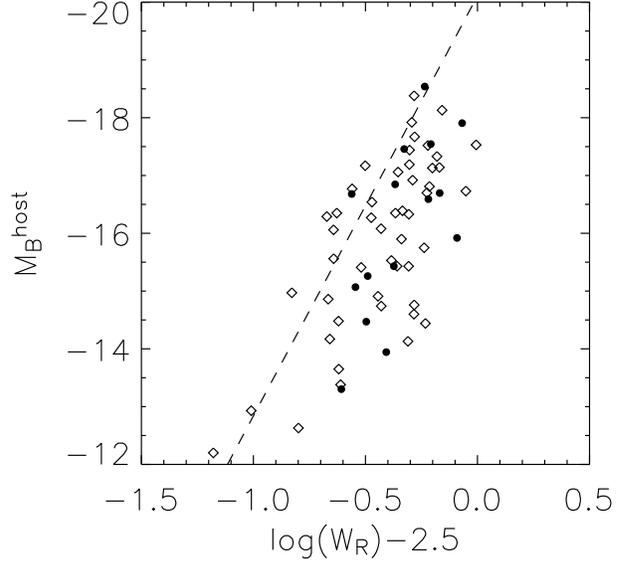}
     \caption{Absolute $B-$band magnitude vs line width for our BCGs (black dots) 
and for isolated dwarf irregulars (open diamonds) from \citet{Vanzee01a}. 
The dashed line indicates the Tully--Fisher relation for spirals derived by \citet{Tully&Pierce}.  
     }
     \label{F9}
   \end{figure}
%
In Fig.~\ref{F8} we present for the BCG hosts, the $L_{\rm B}-\log($v$_{\rm circ})$, 
M$_{\rm s}-\log($v$_{\rm circ})$, and M$_{\rm g+s}-\log($v$_{\rm circ})$ planes for which 
H{\sc i} data are available in the literature (see Table~\ref{T1}). We fitted a straight
 line to each dataset, taking the errors in luminosities and circular velocities into account. 
 In the diagrams we also plotted the TFR, sTFR, and gsTFR fitted by \citet{DeRick07} for 
 a large sample of late-type galaxy discs from various literature sources (including 
 \citet{Mcgaugh00} and \citet{Geha06}). We show its 1$\sigma$ upper 
 and lower limits. The $B$-band luminosities and the corresponding uncertainties for the host
  sample are those derived from our 2D fits (see Table~\ref{T2}). Circular velocities were 
  obtained as $v_{\rm circ}=W_{\rm R}/2$, where $W_{\rm R}$ is the neutral hydrogen line
   width, $W_{\rm 20}$ (see Table~\ref{T1}), after corrections for inclination and random motions \citep{Pier99}
\begin{equation}
W_{\rm R}=(W_{\rm 20}-\Delta W)/\sin(i)
\end{equation}
where the term ${\sin (i)}^{-1}$ corrects for inclination and was obtained following \citet{Stave92}, 
by using the $q$ ($=b/a$) values from the 2D fits. 
The correction due to velocity dispersion, $\Delta W$, was obtained 
following \citet{Bottin83}, under the assumption of a purely Gaussian velocity distribution 
\begin{equation}
\Delta W=3.78 [1.5/(2.25-1.25 \sin^2(i))]^{1/2} \sigma_{z}
\end{equation}
where $\sigma_{z}$ is the velocity dispersion along the $z-$axis. We 
adopted $\sigma_{z}=$5 km s$^{-1}$  \citep{StilIsrael98,Stave92}. Uncertainties for circular 
velocities were estimated from those of $W_{\rm 20}$ ($\sim$5-12 km s$^{-1}$, on average). 

\citet{Bell&deJong} fitted a set of spectrophotometric-disc evolutionary models to a set of 
observed properties of late-type galaxies. Their best models produce a tight correlation 
between the logarithm of the stellar mass-to-light ratio ($M/L$) of the disc and colour.  
We used the colours of the hosts to estimate the {$M/L$} values straightforwardly from these 
correlations. From the several calibrations presented by \citet{Bell&deJong}, we used the one 
obtained by assuming a mass-dependent formation epoch model with bursts and a scaled-down Salpeter 
IMF (see their Table~1). Since these models has negligible dependence with metallicity, the main 
source of uncertainty in the adopted correlations proceeds from the model assumptions, in 
particular the adopted IMF \citep{Bell&deJong}.   
Thus, we computed the stellar masses ($M_{\rm s}$) of the BCG 
underlying stellar population by using $M/L$ and the luminosity of the 2D models. 
Both, $M/L$ and $M_{\rm s}$ are presented in Table~\ref{T4} for the analysed sample.
Moreover, the baryonic mass is calculated as the sum of the stellar mass and the H{\sc i} gas mass 
(from Table~\ref{T4}). 
Combining the different contributions, i.e.\ $B$ luminosity, colours and $M/L$, we estimate a 
 substantial uncertainty in $M_{\rm s}$ and $M_{\rm g+s}$ of up to a factor of 2.  

Figure~\ref{F8}$a$ shows the TFR for late-type spirals and our sample. The relation traced by the 
BCG hosts shows a  slightly steeper slope than those traced by the late-type spiral discs. Our 
linear fit shows that low-luminosity hosts deviate from the extrapolation of the TFR for
 luminous spirals by being fainter than predicted for their circular velocities. We find that 
 8/15 hosts are clearly below the 1$\sigma$ level of the TFR. The observed deviation for  
 late-type dwarf galaxies is not new in the literature. It has already been noticed even when comparing it with
  TFRs of different samples of more luminous systems \citep[see e.g.,][]{StilIsrael98}. 
  \citet{Pier99} points out that the distribution of dwarf galaxies with comparable rotational 
  and random velocities in the $L_{\rm B}-\log($v$_{\rm circ})$ plane is consistent with the 
  TFR for more luminous systems, whereas the rotationally supported dwarfs are underluminous 
  with respect to the extremely late-type galaxies of the same rotational velocity. 

After studying the rotation curves of a sample of 16 rotating dEs, \citet{Vanzee04b} claim that some 
dEs and isolated dIs show remarkable similarities in their kinematic properties. In particular, 
both dwarf types coexist in the same position in the luminosity--line width plane (see their Fig.~6). 
In Fig.~\ref{F9} we reproduce the same plot as presented by \citet{Vanzee04b} but including the host data 
points and excluding the rotating dEs (share the same positions as dIs). Thus, Fig.~\ref{F9} 
shows the TFR derived by \citet{Tully&Pierce} with our data points and those for dIs from \citet{Vanzee00}. 
We see that BCG hosts and dIs (and therefore dEs) are indistinguishable in the luminosity--linewidth plane, 
showing, as in Fig.~\ref{F8}, an overall deviation from the faint end TFR for spiral discs.

In Fig.~\ref{F8}$b$ we show the sTFR for late-type spirals and our sample data points in the 
 M$_{\rm s}-\log(v_{\rm circ})$ plane. While the sTFR reduces its scatter for late-type discs, the 
 opposite behaviour is observed for BCG hosts (compared with Fig.~\ref{F8}$a$). Our weighted linear 
 fit to our sample data points, although more uncertain, exhibits a slope much steeper than in Fig.~\ref{F8}$a$.
 
In contrast, in Fig.~\ref{F8}$c$ our data points show a smaller scatter compared with 
Figs.~\ref{F8}$a$ and \ref{F8}$b$. Furthermore, the slope of the gsTFR traced by the hosts 
is much steeper than the slope of the extrapolated gsTFR for late-type spirals. From this, {\em 
we expect low-luminosity hosts to be less massive than predicted for their circular velocities}. 
On average, $\sim$50\% of the BCG hosts are undermassive by a factor of $\sim$0.4 dex, suggesting 
that they could have lost part of their baryons. Similar results have been obtained for a sample of 
dE by \citet{DeRick07}.  

\begin{figure} 
\centering
 \includegraphics[angle=90,width=7.5cm]{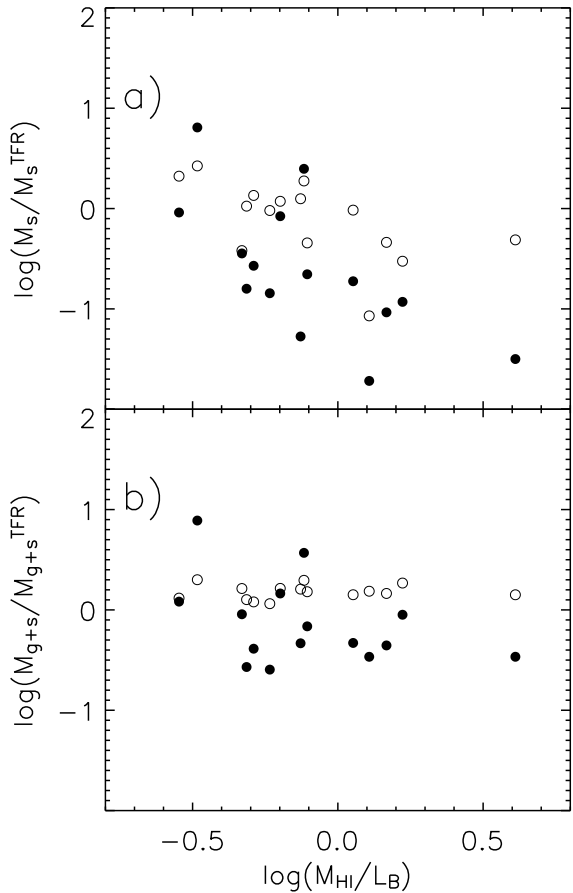}\\
      \caption{Deviations from the stellar and baryonic Tully--Fisher relations: {\em a)} the ratio 
      between the host stellar mass and the expected stellar mass from the sTFR, and {\em b)} the 
      ratio between the host baryonic mass and the expected baryonic mass from the gsTFR as functions 
      of the H{\sc i} mass to $B$-band luminosity ratio (see text for details). 
      }
   \label{F10}
   \end{figure}

\citet{Warren07} have recently found that dI galaxies with generally high M$_{\rm HI}/L_{\rm B}$ 
ratios follow the same trend in the baryonic TFR as defined by lower M$_{\rm HI}/L_{\rm B}$ giant
 galaxies, but are too underluminous for their rotation to follow the trend in an sTFR, {\em suggesting 
 that the baryonic mass of  dwarf galaxies tends to be normal but that these have failed to produce 
 sufficient numbers of stars}.   
In Figs~\ref{F10}$a$, and~\ref{F10}$b$ we have plotted with filled circles the host residuals from 
the stellar and baryonic TFR versus the H{\sc i} mass to $B$-band luminosity ratio. We also calculated 
the stellar mass and baryonic mass of the hosts, which should correspond to them if they follow the same 
TFR and sTFR as the late-type discs of \citet{DeRick07}. We have overplotted these masses with open circles in 
Figs.~\ref{F10}$a$,~\ref{F10}$b$. 
The stellar mass (M$_{\rm s}$) deviation from the sTFR correlates with the distance-independent ratio
 M$_{\rm HI}/L_{\rm B}$ (Fig.~\ref{F10}$a$), even if our BCGs should follow the TFR for late-type discs 
 (open circles). In contrast, the H{\sc i} gas$+$ stellar mass (M$_{\rm g+s}$) deviations from 
 the gsTFR do not show such a correlation (Fig.~\ref{F10}$b$). Even if BCGs should have enough  
 stars (i.e.\ if they follow the normal sTFR), their baryonic mass should tend to those 
 predicted by the gsTFR for late-type discs. Overall, this means that BCGs tend to a normal baryonic 
 content and that those galaxies that fall below the gsTFR have not lost a significant amount of their 
 gas by galactic winds or environmental effects. On the contrary, these deviations from the normal Tully--Fisher 
 relations show that the overall host stellar mass is too low to reach the trend of spiral discs. 
 In particular, galaxies with higher gas fractions, which are those with blue colours (see Fig.~\ref{F6}$c$), 
 are undermassive for their circular velocities. Our result agrees with those of \citet{Warren07}, 
 and suggests that gas-rich BCGs tend to be inefficient at forming stars, such as dIs.    
%
%
%

All the above shows that, on average, dynamical and structural connections between BCGs, dIs and 
dEs cannot be ruled out and suggests that the BCG ``zoo'' must be differentiated by the structure 
and colours of the underlying stellar population when exploring the possible evolutionary scenarios 
between the different ``families'' of dwarf galaxies.   

\section{Summary}
We have analysed the underlying stellar host of a sample of 28 BCG galaxies by fitting their deep optical 
images. The host was modelled by a S\'ersic law, masking the starburst contribution to the broad band 
images and by accurately subtracting  the sky background in the consistent way inherited from  
Paper~I. Only the best 2D S\'ersic models were accepted for those galaxies that fulfilled the consistency 
checks. Based on this, we reduced our full sample to 20 galaxies that, according to 
their absolute $B$-band magnitudes, may be considered as dwarf systems. Galaxies for which 
their S\'ersic models were discarded ($\sim$28\%) have been reported in the literature as galaxies with 
clear signs of interactions/mergers or grouping. 

Two-dimensional S\'ersic models show low S\'ersic indices, $0.5 \la n \la 2$, for all galaxies but one. 
In addition, while 
effective radii show a positive correlation with luminosity, the extinction-corrected central surface 
brightness shows a large scatter, with an average value $<$$\mu_{\rm 0,B}$$> =$ 20.28$\pm$1.44 mag 
arcsec$^{-2}$. S\'ersic models show a wide range of colours ($<$$B-R$$> \sim$ 1), while no correlations 
between colours and the fitted S\'ersic parameters, i.e.\ $n$, $r_{\rm e}$, and $\mu_{\rm e}$, were found. 
However, the sample could be split into different groups according to their $B-R$ colours. A ``red'' group 
showing $B-R$ colours redder ($<$$B-R$$> =$ 1.29$\pm$ 0.10) than the median for the whole sample ($<$$B-R$$> =$ 0.97) 
can be distinguished among the more luminous systems, whereas a ``blue'' group ($<$$B-R$$> =$ 0.66$\pm$0.07) 
consists, on average, of fainter objects. 

The sizes and luminosities of both the host and the starburst components also show strong correlations. 
More extended and luminous hosts harbour more extended and luminous starbursts. In contrast, little or 
no correlation was found between $L_{\rm burst}/L_{\rm host}$ and the host luminosity, indicating that 
the relative strength of the bursts does not depend on the galaxy mass. On the other hand, the H{\sc i} gas 
mass correlates with the host luminosity, showing that galaxies with larger gas reservoirs are expected to
 be optically more luminous and also more extended. The above relations suggest that star formation 
 in this type of galaxies is probably self-regulated and that the star formation could be proceeding in a 
positive feedback regime. 
Moreover, galaxies with lower M$_{\rm HI}/L_{\rm B}$ show a tendency to have redder hosts. 
 
We compared the host structural properties with those of other galaxy types. The S\'ersic parameters 
of the BCG hosts and their global trends with  luminosity are similar to those shown by dIs and some dEs. 
In particular, most compact and redder hosts are structurally similar to some dEs. Nevertheless, the overall 
sample of hosts is about twice as compact and shows a central surface brightness about two magnitudes brighter than dIs.  

We explored the BCG hosts in the TFR context and compared their position to those of late-type
 discs, taking the host $B$-band luminosities into account, as well as the stellar and H{\sc i} masses. 
The BCG hosts, isolated dIs, and rotating dEs turned out to be indistinguishable in the $B$-band luminosity--line width 
plane. We found that about 50--60\% of the galaxies are located below the TFR followed by late-type spiral 
discs. A larger scatter was obtained in the stellar mass--circular velocity plane, from which we have induced 
that host are clearly undermassive for their circular velocity if they follow the sTFR as late-type discs. The 
deviations from the sTFR correlate with the distance-independent H{\sc i} mass-to-luminosity ratio, showing
 that hosts with higher M$_{\rm HI}/L_{\rm B}$ have lower stellar masses than expected for the sTFR of more 
 luminous objects. In the gsTFR, the baryonic mass is slightly lower than predicted for late-type discs for 
approximately half of the galaxies of a given circular velocity. Nevertheless, these differences disappear if 
  we consider the stellar mass for the host predicted by the sTFR for spirals at a given v$_{\rm circ}$, 
  indicating that these deviations are a product of an insufficient amount of stellar mass rather than a result 
  of mass loss by starburst-driven galactic winds. Furthermore, deviations from the gsTFR do not correlate with 
  M$_{\rm HI}/L_{\rm B}$, indicating that the overall baryonic mass in these systems tends to be normal, but 
  it is deficient in stars, in a similar way to what has recently been observed for dI galaxies.

\begin{acknowledgements}
 The authors are sincerely grateful to the referee for the very detailed comments, suggestions, 
and time, all of which helped us to improve the manuscript.
Thanks are given to N. Caon for his strong participation in the first stages of this project and 
to Jorge S\'anchez Almeida and Rub\'en S\'anchez-Janssen for their very helpful comments. 
We are also grateful to Terry Mahoney for revising the English. 
This paper is based on observations made at the Nordic Optical Telescope (NOT), at the Spanish Observatory 
del Roque de los Muchachos of the Instituto de Astrof\'{\i}sica  de Canarias and on observations taken at 
the German/Spanish Calar Alto Observatory. 
This work was partially funded by the Spanish DGCyT, grants 
AYA2007-67965-C03-01 and AYA2007-67965-C03-02. 
This research was also partially funded by the Spanish MEC under the Consolider-Ingenio 2010
Programme grant CSD2006-00070: First Science with the GTC
(http://www.iac.es/consolider-ingenio-gtc/)
This research made use of the NASA/IPAC Extragalactic Database
 (NED) which is operated by the Jet Propulsion Laboratory, California Institute of Technology, under 
 contract with the National Aeronautics and Space Administration. 
\end{acknowledgements}


\footnotesize{
\begin{longtable}{c c c c c c c c c c c c c c}
\caption{\label{T2} Parameters for the underlying stellar host derived from the 2D \mbox{S\'ersic} fit}\\
\noalign{\smallskip}
\hline\hline \\ 
Galaxy & Filter & $PA$ & $q$ & $c$ & \mbox{$R_{\rm tran}$} & $n$ & \mbox{$r_{\rm e}$} & m$_{\rm host}$ &
 $M_{\rm host}$ & $\mu_{\rm e}$ & $\mu_{\rm 0}$ & $\mu_{\rm tran}$ &  $Q$ \\
\noalign{\smallskip}
   (1)  &   (2)  &   (3)  &   (4)   &   (5)   &   (6)  & (7) & (8) &  (9)  &  (10) & (11)  & (12)&  (13)& (14) \\[3pt] 
\noalign{\smallskip}
\hline
\noalign{\smallskip}
\endfirsthead
\caption{continued.}\\
\noalign{\smallskip}
\hline\hline\\
Galaxy & Filter & $PA$ & $q$ & $c$ & \mbox{$R_{\rm tran}$} & $n$ & \mbox{$r_{\rm e}$} & m$_{\rm host}$ &
 $M_{\rm host}$ & $\mu_{\rm e}$ & $\mu_{\rm 0}$ &  $\mu_{\rm tran}$ & $Q$ \\
\noalign{\smallskip}
   (1)  &   (2)  &   (3)  &   (4)   &   (5)   &   (6)  & (7) & (8) &  (9)  &  (10) & (11)  & (12)&  (13)& (14) \\[3pt] 
\noalign{\smallskip}
\hline
\noalign{\smallskip}
\endhead
\hline
\endfoot

Cam0341-4045&$B$&$-$80.9&0.73&   1.19&1.59&0.74$\pm$0.20&1.17$\pm$0.08&17.45$\pm$0.12&$-$16.32&21.90&20.65&23.46& 1 \\[3pt]	  
            &$V$&$-$73.9&0.77&   0.44&1.59&0.97$\pm$0.21&1.31$\pm$0.08&17.08$\pm$0.14&$-$16.69&23.26&21.51&23.26& 1 \\[3pt]	  
            &$R$&$-$83.5&0.84&$-$0.46&1.59&0.99$\pm$0.27&1.53$\pm$0.22&16.49$\pm$0.21&$-$17.28&23.32&21.53&22.79& 1 \\[3pt]  
Haro~1      &$B$&$-$15.2&0.97&   0.05&3.33&1.44$\pm$0.21&1.65$\pm$0.04&13.08$\pm$0.25&$-$20.51&19.99&17.22&21.73& 2 \\[3pt]	  
	    &$V$&   80.0&0.99&   0.18&4.37&1.39$\pm$0.20&1.69$\pm$0.22&12.45$\pm$0.22&$-$21.14&19.41&16.76&22.03& 2 \\[3pt]	  
	    &$R$&$-$44.6&0.99&$-$0.15&4.37&1.37$\pm$0.25&1.70$\pm$0.04&12.70$\pm$0.28&$-$20.89&19.67&17.05&22.26& 2 \\[3pt]	  
	    &$I$&   83.4&0.97&$-$0.04&4.64&1.35$\pm$0.25&1.81$\pm$0.23&11.68$\pm$0.35&$-$21.90&18.76&16.18&21.37& 2 \\[3pt] 
Haro~15     &$B$& 10.7  &0.72&$-$0.05&2.77&0.85$\pm$0.12&1.57$\pm$0.29&14.31$\pm$0.24&$-$20.32&19.49&18.01&20.92& 2 \\[3pt] 
	    &$V$& 16.8  &0.81&   0.00&3.54&1.14$\pm$0.32&1.78$\pm$0.20&14.60$\pm$0.34&$-$20.02&20.33&18.21&22.09& 2 \\[3pt]	  
	    &$R$& 11.2  &0.78&   0.00&2.81&1.02$\pm$0.37&1.33$\pm$0.45&13.77$\pm$0.44&$-$20.86&18.77&16.92&20.78& 2 \\[3pt]  
Mrk~5       &$B$& 13.1  &0.76&   0.10&0.74&0.99$\pm$0.15&0.65$\pm$0.08&15.66$\pm$0.20&$-$15.07&22.96&21.18&23.22& 1 \\[3pt] 
            &$V$& 13.2  &0.75&   0.07&0.74&1.05$\pm$0.14&0.64$\pm$0.03&15.26$\pm$0.08&$-$15.46&22.56&20.64&22.84& 1 \\[3pt] 
            &$R$& 12.5  &0.78&$-$0.01&0.71&1.05$\pm$0.25&0.64$\pm$0.02&14.98$\pm$0.08&$-$15.74&22.32&20.41&22.54& 1 \\[3pt]  
Mrk~33      &$B$&$-$63.4&0.64&   0.23&1.72&2.88$\pm$0.71&1.71$\pm$0.40&14.20$\pm$0.12&$-$17.54&22.93&17.03&22.94& 1 \\[3pt]  
	    &$V$&$-$64.1&0.61&   0.31&2.00&1.88$\pm$0.60&1.74$\pm$0.50&13.84$\pm$0.23&$-$17.90&22.33&18.61&22.62& 1 \\[3pt]	  
	    &$R$&$-$64.3&0.62&   0.32&1.71&2.24$\pm$0.20&1.65$\pm$0.37&12.95$\pm$0.10&$-$18.79&21.43&16.93&21.52& 1 \\[3pt]  
Mrk~35      &$B$& 74.0  &0.72&$-$0.02&1.57&0.99$\pm$0.18&1.16$\pm$0.12&14.12$\pm$0.18&$-$16.85&22.38&20.59&23.03& 1 \\[3pt] 
            &$V$& 77.3  &0.71&   0.03&1.47&1.01$\pm$0.27&1.12$\pm$0.10&13.39$\pm$0.15&$-$17.58&21.57&19.74&22.13& 1 \\[3pt] 
            &$R$& 77.7  &0.70&$-$0.01&1.47&0.97$\pm$0.16&1.13$\pm$0.08&13.09$\pm$0.12&$-$17.87&21.26&19.51&21.80& 1 \\[3pt]  
Mrk~36      &$B$&$-$25.9&0.47&$-$0.00&0.51&1.00$\pm$0.04&0.44$\pm$0.04&16.79$\pm$0.15&$-$13.30&23.39&21.57&23.66& 1 \\[3pt] 
            &$V$&$-$24.8&0.48&$-$0.12&0.52&1.07$\pm$0.13&0.44$\pm$0.03&16.24$\pm$0.25&$-$13.85&22.85&20.92&23.21& 1 \\[3pt] 
            &$R$&$-$23.8&0.49&$-$0.10&0.52&1.05$\pm$0.08&0.45$\pm$0.03&15.87$\pm$0.10&$-$14.22&22.54&20.62&22.85& 1 \\[3pt]  
Mrk~86      &$B$&$-$21.2&0.91&   0.00&1.37&0.94$\pm$0.34&1.01$\pm$0.19&12.87$\pm$0.27&$-$16.68&22.48&20.80&23.12& 1 \\[3pt] 
            &$V$&$-$23.7&0.92&   0.01&1.36&1.00$\pm$0.32&1.03$\pm$0.15&12.13$\pm$0.20&$-$17.42&21.83&20.02&22.42& 1 \\[3pt] 
            &$R$&$-$21.3&0.90&   0.00&1.36&1.02$\pm$0.26&1.00$\pm$0.27&11.59$\pm$0.30&$-$17.96&21.21&19.36&21.87& 1 \\[3pt] 
Mrk~297     &$B$&$-$86.3&0.71&$-$0.18&6.31&0.84$\pm$0.15&2.92$\pm$1.00&14.27$\pm$0.31&$-$19.80&21.35&19.88&23.55& 2 \\[3pt]	  
	    &$V$&$-$85.9&0.73&$-$0.17&5.68&0.65$\pm$0.29&3.82$\pm$1.07&14.20$\pm$0.54&$-$19.87&21.78&20.73&22.66& 2 \\[3pt]  
Mrk~314     &$B$& 14.8  &0.77&$-$0.46&1.51&1.01$\pm$0.17&1.34$\pm$0.11&14.40$\pm$0.11&$-$17.91&21.71&19.88&21.96& 1 \\[3pt]
Mrk~324     &$B$&3.75	&0.82&$-$0.15&0.51&1.12$\pm$0.23&0.40$\pm$0.02&15.83$\pm$0.13&$-$15.92&21.21&19.14&21.70& 1 \\[3pt]	  
	    &$V$&$-$14.4&0.88&   0.11&0.51&1.12$\pm$0.21&0.41$\pm$0.05&15.11$\pm$0.22&$-$16.64&20.61&18.54&21.06& 1 \\[3pt]  
	    &$R$&$-$17.5&0.88&   0.03&0.51&1.11$\pm$0.26&0.44$\pm$0.04&14.75$\pm$0.18&$-$17.01&20.41&18.36&20.68& 1 \\[3pt] 
	    &$I$&$-$17.5&0.88&   0.03&0.51&1.02$\pm$0.24&0.42$\pm$0.04&14.38$\pm$0.23&$-$17.38&19.89&18.02&20.27& 1 \\[3pt] 
Mrk~370     &$B$&$-$14.8&0.79&   0.09&1.00&1.04$\pm$0.08&1.23$\pm$0.05&13.48$\pm$0.13&$-$16.70&23.18&21.28&22.85& 1 \\[3pt] 
            &$V$&$-$18.4&0.81&   0.09&0.92&1.00$\pm$0.10&1.22$\pm$0.07&12.93$\pm$0.10&$-$17.24&22.53&20.72&22.10& 1 \\[3pt] 
            &$R$&$-$14.0&0.80&   0.08&0.88&1.03$\pm$0.14&1.26$\pm$0.07&12.46$\pm$0.10&$-$17.71&22.07&20.19&21.52& 1 \\[3pt]  
Mrk~401	    &$B$&   35.1&0.97&$-$0.01&1.69&0.49$\pm$0.22&2.28$\pm$0.10&14.45$\pm$0.17&$-$17.46&23.27&22.56&22.95& 1 \\[3pt]	  
	    &$R$&$-$62.5&1.00&$-$0.12&1.80&0.50$\pm$0.17&2.11$\pm$0.11&13.81$\pm$0.18&$-$18.10&22.50&21.78&22.31& 1 \\[3pt]  
Mrk~527     &$B$&$-$65.7&0.68&   0.23&1.68&1.04$\pm$0.30&2.99$\pm$0.35&14.10$\pm$0.20&$-$19.29&21.96&20.06&21.15& 1 \\[3pt]	  
	    &$V$&$-$64.5&0.79&   0.37&1.95&1.01$\pm$0.26&2.81$\pm$0.28&13.45$\pm$0.20&$-$19.94&21.33&19.49&20.77& 1 \\[3pt]	  
	    &$I$&$-$62.8&0.81&   0.31&3.22&1.22$\pm$0.32&2.73$\pm$0.30&12.13$\pm$0.19&$-$21.26&20.05&17.76&20.39& 1 \\[3pt]  
Mrk~600     &$B$&$-$45.0&0.53&   0.52&0.44&1.56$\pm$0.16&0.36$\pm$0.03&15.28$\pm$0.15&$-$15.26&21.30&18.27&21.76& 1 \\[3pt]	  
	    &$V$&$-$44.1&0.55&   0.35&0.42&1.44$\pm$0.15&0.39$\pm$0.04&14.98$\pm$0.13&$-$15.55&21.20&18.43&21.38& 1 \\[3pt]	  
	    &$R$&$-$44.1&0.53&   0.48&0.42&1.55$\pm$0.12&0.40$\pm$0.02&14.74$\pm$0.09&$-$15.79&20.98&17.98&21.14& 1 \\[3pt] 
Mrk~1089    &$B$&   39.4&0.70&   0.00&3.82&0.77$\pm$0.31&4.09$\pm$0.63&14.64$\pm$0.55&$-$18.97&22.85&21.54&22.75& 2 \\[3pt]	  
	    &$V$&   44.2&0.68&   0.00&3.82&0.67$\pm$0.28&3.85$\pm$0.58&14.58$\pm$0.30&$-$19.03&22.58&21.48&22.57& 2 \\[3pt]	  
	    &$R$&   51.1&0.64&   0.00&3.82&0.92$\pm$0.33&2.83$\pm$1.42&13.73$\pm$0.65&$-$19.87&21.13&19.49&21.77& 2 \\[3pt]  
Mrk~1090    &$B$& $-$6.7&1.00&$-$0.11&1.66&0.84$\pm$0.17&1.11$\pm$0.10&14.77$\pm$0.15&$-$18.81&20.61&19.15&21.51& 2 \\[3pt]	  
	    &$V$& $-$6.5&1.00&$-$0.12&1.81&0.87$\pm$0.14&1.18$\pm$0.08&14.69$\pm$0.15&$-$18.89&20.68&19.15&21.65& 2 \\[3pt]	  
	    &$R$&$-$10.7&1.00&$-$0.12&1.66&0.90$\pm$0.17&1.17$\pm$0.13&14.29$\pm$0.16&$-$19.29&20.26&18.67&21.04& 2 \\[3pt]	  
	    &$I$&    5.7&0.96&$-$0.19&1.66&0.83$\pm$0.08&1.09$\pm$0.08&13.80$\pm$0.15&$-$19.78&19.55&18.11&20.50& 2 \\[3pt] 
PHL~293B    &$B$&$-$26.3&0.89&   0.24&0.24&0.59$\pm$0.14&0.32$\pm$0.05&17.71$\pm$0.24&$-$13.94&22.49&21.57&22.17& 1 \\[3pt]	  
	    &$V$&$-$26.6&0.91&   0.06&0.40&0.53$\pm$0.17&0.35$\pm$0.04&17.13$\pm$0.28&$-$14.52&22.07&21.28&22.37& 1 \\[3pt]	  
	    &$R$&$-$25.5&0.92&   0.07&0.30&0.58$\pm$0.15&0.39$\pm$0.05&16.99$\pm$0.15&$-$14.66&22.24&21.34&21.94& 1 \\[3pt]  
Tol~0127-397&$B$&$-$53.8&0.69&   0.23&1.35&1.11$\pm$0.19&1.19$\pm$0.23&17.35$\pm$0.19&$-$16.57&22.71&20.66&22.97& 1 \\[3pt]
            &$V$&$-$50.5&0.67&   0.08&1.40&0.94$\pm$0.15&1.37$\pm$0.20&16.03$\pm$0.20&$-$17.90&21.58&19.90&21.63& 1 \\[3pt] 
            &$R$&$-$50.8&0.66&   0.08&1.55&1.02$\pm$0.24&1.40$\pm$0.06&15.92$\pm$0.21&$-$18.01&21.56&19.70&21.75& 1 \\[3pt]  
UM~417      &$B$&$-$19.2&0.50&$-$0.22&0.39&0.68$\pm$0.12&0.71$\pm$0.10&18.36$\pm$0.16&$-$14.37&23.29&22.18&22.62& 1 \\[3pt]	  
	    &$V$&$-$17.5&0.52&$-$0.34&0.39&0.49$\pm$0.16&0.70$\pm$0.09&18.14$\pm$0.19&$-$14.58&22.90&22.20&22.41& 1 \\[3pt]    
UM~462      &$B$&    9.8&0.77&   0.00&0.92&1.80$\pm$0.34&0.51$\pm$0.08&15.26$\pm$0.28&$-$15.49&22.31&18.76&23.70& 1 \\[3pt]	  
	    &$V$&    9.0&0.76&   0.00&0.80&1.50$\pm$0.30&0.50$\pm$0.05&15.07$\pm$0.26&$-$15.68&21.98&19.09&23.04& 1 \\[3pt]	  
	    &$R$&    9.3&0.70&   0.00&0.79&1.85$\pm$0.36&0.67$\pm$0.11&14.52$\pm$0.13&$-$16.23&22.09&18.43&22.43& 1 \\[3pt]  
I~Zw~123    &$B$&$-$51.1&0.92&$-$0.26&0.43&1.58$\pm$0.20&0.28$\pm$0.04&16.02$\pm$0.20&$-$14.47&22.15&19.08&23.14& 1 \\[3pt] 
            &$V$&$-$52.9&0.89&$-$0.20&0.43&2.20$\pm$0.17&0.25$\pm$0.02&15.41$\pm$0.14&$-$15.07&21.43&17.01&22.69& 1 \\[3pt]  
II~Zw~33    &$B$&   70.2&0.88&$-$0.68&2.05&1.02$\pm$0.38&1.40$\pm$0.28&13.23$\pm$0.42&$-$19.58&22.46&20.60&21.14& 2\\[3pt]  
	    &$V$&$-$22.4&1.00&$-$0.65&1.90&0.90$\pm$0.40&1.43$\pm$0.24&12.22$\pm$0.46&$-$20.58&20.52&18.93&20.01& 2\\[3pt]	       
	    &$I$&   23.5&1.00&   1.69&1.90&0.80$\pm$0.31&1.65$\pm$0.31&11.68$\pm$0.51&$-$21.13&19.64&18.26&19.40& 2\\[3pt]  
II~Zw~40    &$B$&$-$41.6&0.70&   0.00&0.92&0.92$\pm$0.21&0.91$\pm$0.16&12.57$\pm$0.15&$-$17.36&21.28&19.64&21.30& 2\\[3pt]	       
	    &$V$&$-$44.5&0.61&   0.00&0.92&1.29$\pm$0.30&1.05$\pm$0.15&12.41$\pm$0.16&$-$17.52&21.44&19.00&21.20& 2\\[3pt]	       
	    &$R$&$-$55.1&0.72&   0.00&0.70&1.15$\pm$0.14&1.25$\pm$0.21&11.85$\pm$0.17&$-$18.08&21.37&19.24&20.38& 2\\[3pt]
II~Zw~70    &$B$& 56.2  &0.30&$-$0.71&1.08&1.58$\pm$0.28&0.85$\pm$0.14&15.98$\pm$0.26&$-$15.43&21.43&18.36&22.91& 1 \\[3pt]	  
	    &$V$& 56.4  &0.27&$-$0.69&1.10&1.07$\pm$0.20&0.86$\pm$0.19&15.69$\pm$0.28&$-$15.72&22.13&20.17&22.34& 1 \\[3pt]	  
	    &$R$& 55.8  &0.29&$-$0.58&1.02&1.00$\pm$0.18&0.91$\pm$0.06&15.28$\pm$0.23&$-$16.13&22.00&20.20&21.83& 1 \\[3pt]  
II~Zw~71    &$B$& 40.1  &1.00&$-$0.84&1.45&1.26$\pm$0.11&1.35$\pm$0.06&14.83$\pm$0.23&$-$16.59&23.37&21.00&23.58& 1 \\[3pt]	  
	    &$V$& 39.0  &1.00&$-$0.79&1.45&1.31$\pm$0.15&1.27$\pm$0.07&14.87$\pm$0.20&$-$17.16&23.47&20.98&23.03& 1 \\[3pt]	  
            &$R$& 39.1  &0.90&$-$0.78&1.45&1.34$\pm$0.16&1.15$\pm$0.08&14.31$\pm$0.22&$-$17.57&22.66&20.11&22.52& 1 \\[3pt]   
III~Zw~102  &$B$&   40.0&0.93&   0.00&2.45&0.51$\pm$0.15&2.32$\pm$0.16&13.24$\pm$0.18&$-$18.54&22.19&21.45&22.28& 1 \\[3pt]	  
	    &$V$&   40.5&0.95&   0.00&2.82&0.50$\pm$0.13&2.41$\pm$0.11&12.48$\pm$0.15&$-$19.31&21.53&20.81&21.80& 1 \\[3pt]	  
	    &$R$&   50.5&0.93&$-$0.03&2.45&0.56$\pm$0.19&2.25$\pm$0.15&12.03$\pm$0.15&$-$19.75&20.95&20.09&21.10& 1 \\[3pt]  
III~Zw~107  &$B$&    4.7&0.64&   0.43&3.97&1.63$\pm$0.65&2.18$\pm$0.76&15.26$\pm$0.22&$-$19.20&21.50&18.32&22.93& 2\\[3pt]	       
	    &$V$&    3.2&0.54&   0.94&3.97&1.65$\pm$0.53&2.37$\pm$0.61&14.68$\pm$0.18&$-$19.78&20.93&17.71&22.12& 2\\[3pt]	       
	    &$R$&    5.8&0.52&   1.29&3.97&1.67$\pm$0.65&2.41$\pm$0.64&14.88$\pm$0.21&$-$19.58&21.13&17.87&22.27& 2\\[3pt]  
\noalign{\smallskip}
\hline
\hline
\end{longtable}
}
\begin{list}{}{}
\item Notes.$-$ Columns: 
(3) position angle; 
(4) minor-to-major axis ratio ($b/a$); 
(5) boxy/discy shape parameter;  
(6) transition radius, in kiloparsecs;
(7) \mbox{S\'ersic} shape parameter; 
(8) effective radius, in kiloparsecs; 
(9) total apparent magnitude of the host component; 
(10) absolute magnitude of the host component; 
(11) effective surface brightness of the host component, in mag arcsec$^{2}$; 
(12) central surface brightness of the host component, in mag arcsec$^{-2}$; 
(13) surface brightness at the transition radius, in mag arcsec$^{-2}$;
(14) quality index : 1={\em good fit}, 2={\em poor fit}.
 Columns (9) -- (13) were corrected by Galactic extinction following \citet{schlegel98}  \\
\end{list}
%
%

\Online

\begin{figure*}
%
\includegraphics[width=0.98\textwidth,angle=0]{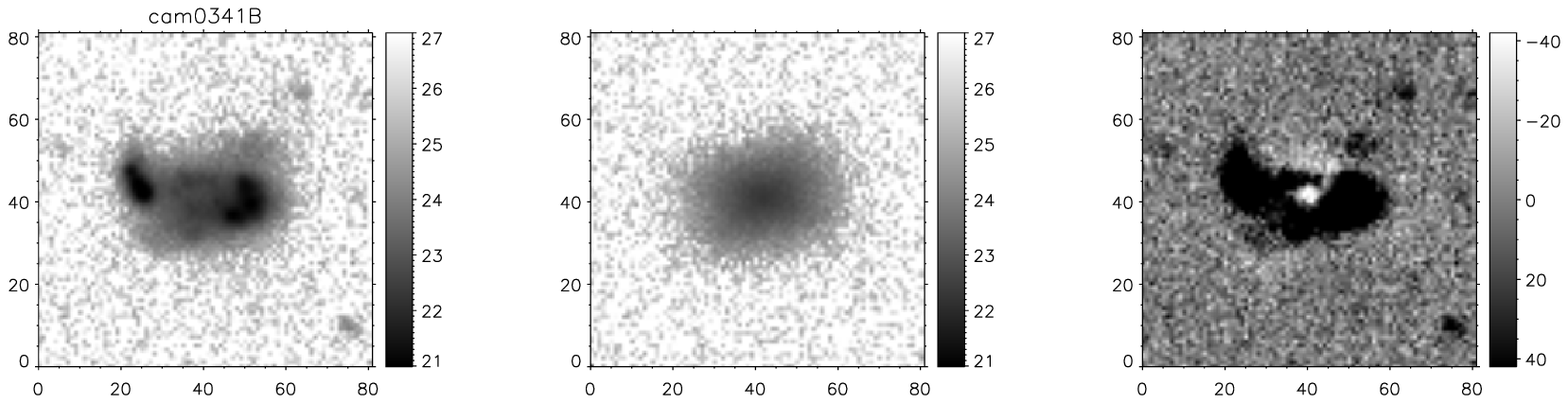}\\[7pt] 
\includegraphics[width=0.98\textwidth,angle=0]{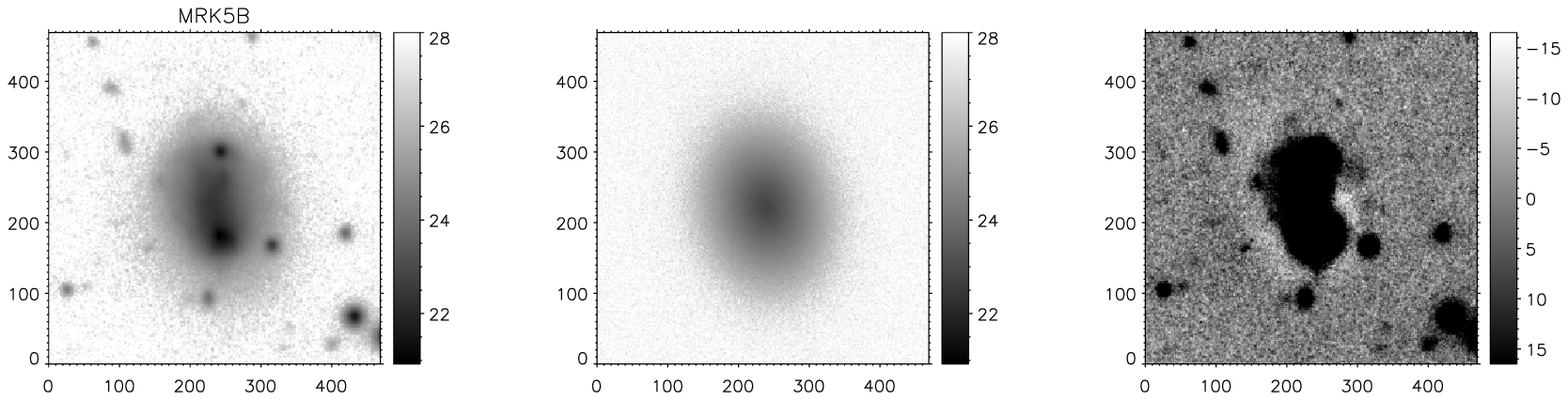}\\[7pt]
\includegraphics[width=0.98\textwidth,angle=0]{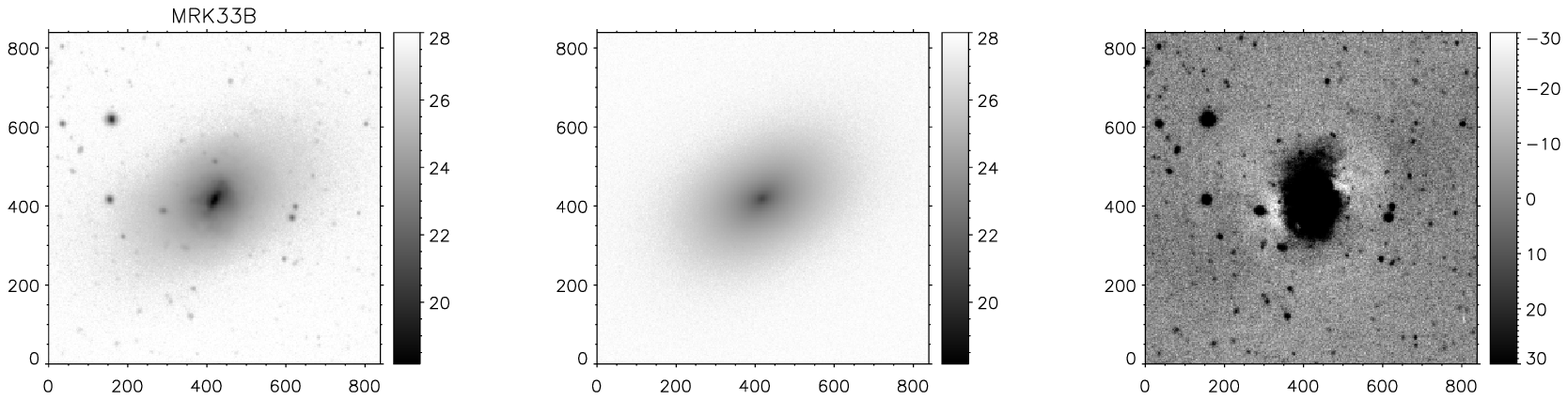}\\[7pt]
\includegraphics[width=0.98\textwidth,angle=0]{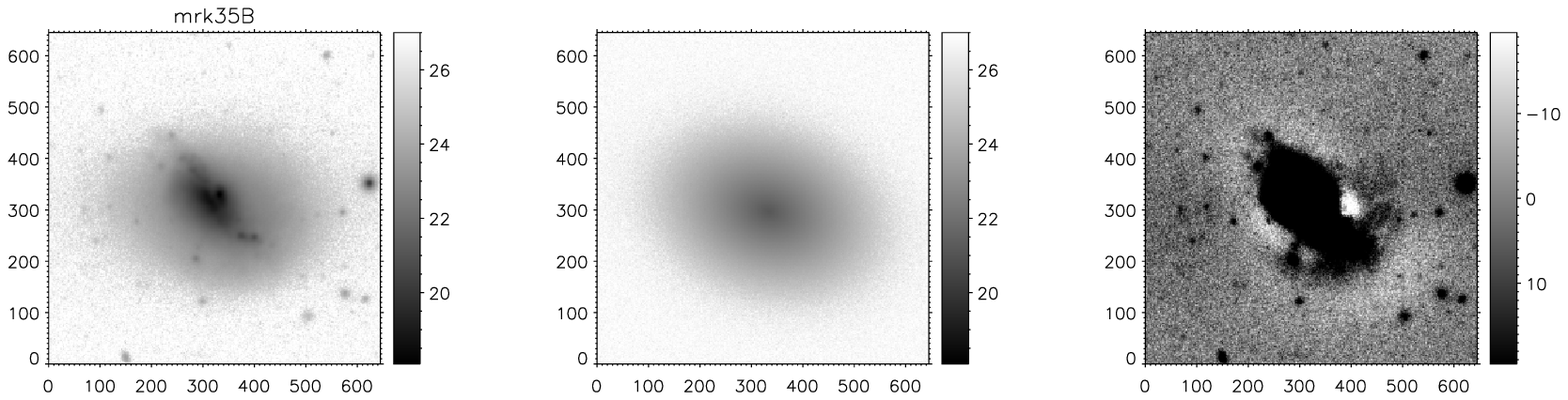}\\[7pt] 
\caption[]{\label{Ap1} Atlas of galaxy images I: Galaxies with $Q=1$. \footnotesize{Broad-band galaxy images (left), the best 2D \mbox{S\'ersic} model (centre) and the residual image (right) of the 20 galaxies fitted with quality index $Q=1$ in the full sample of 28 BCGs. Broad-band images and 2D \mbox{S\'ersic} models are showed in surface brightness grey scale (magnitude persquare arcseconds). Realistic Poissonian noise are included in the 2D models. Residual images, i.e.\ galaxy$-$model, are displayed in an inverted linear grey scale (counts) which covers the range $\pm 3\sigma$, where $\sigma$ is the mean $rms$ of the sky background. Thus, darkest regions are the most luminous regions (mainly the starburst) which have been saturated to enhance residual features in the outer fitted regions, e.g.\ the host galaxy region. In all cases, horizontal and vertical axis are in pixels whereas the orientation is north (up)~--~east (left).}}
\end{figure*}

\begin{figure*}
\ContinuedFloat
\includegraphics[width=0.98\textwidth,angle=0]{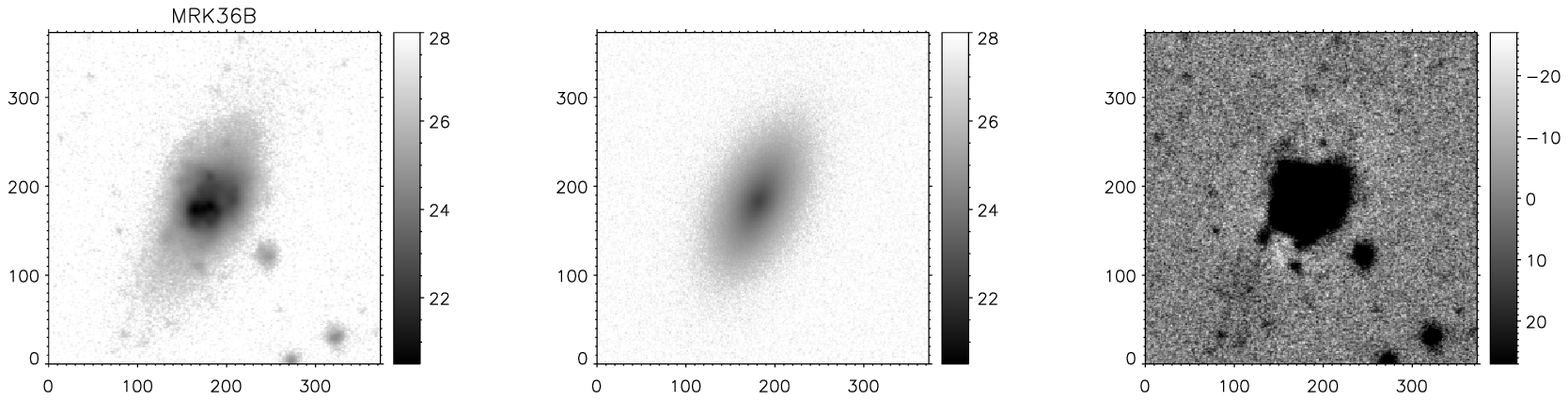}\\[7pt] 
\includegraphics[width=0.98\textwidth,angle=0]{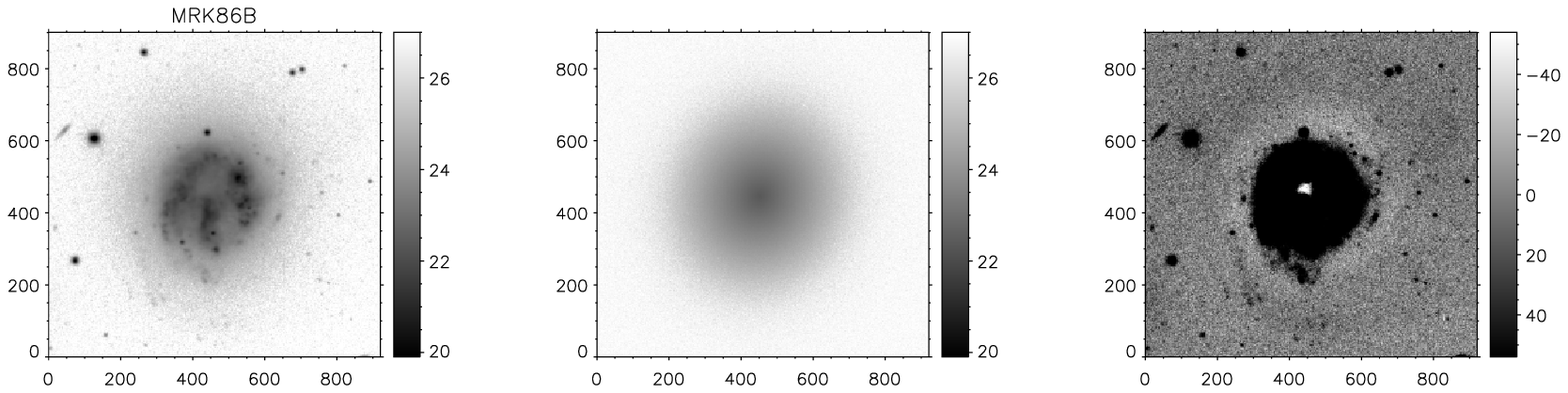}\\[7pt] 
\includegraphics[width=0.98\textwidth,angle=0]{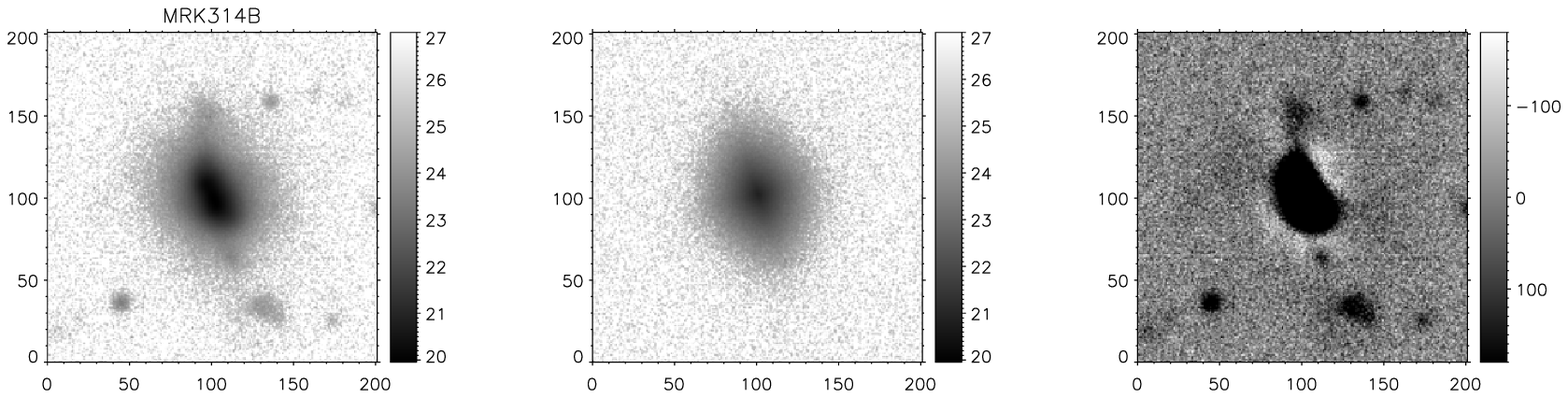}\\[7pt] 
\includegraphics[width=0.98\textwidth,angle=0]{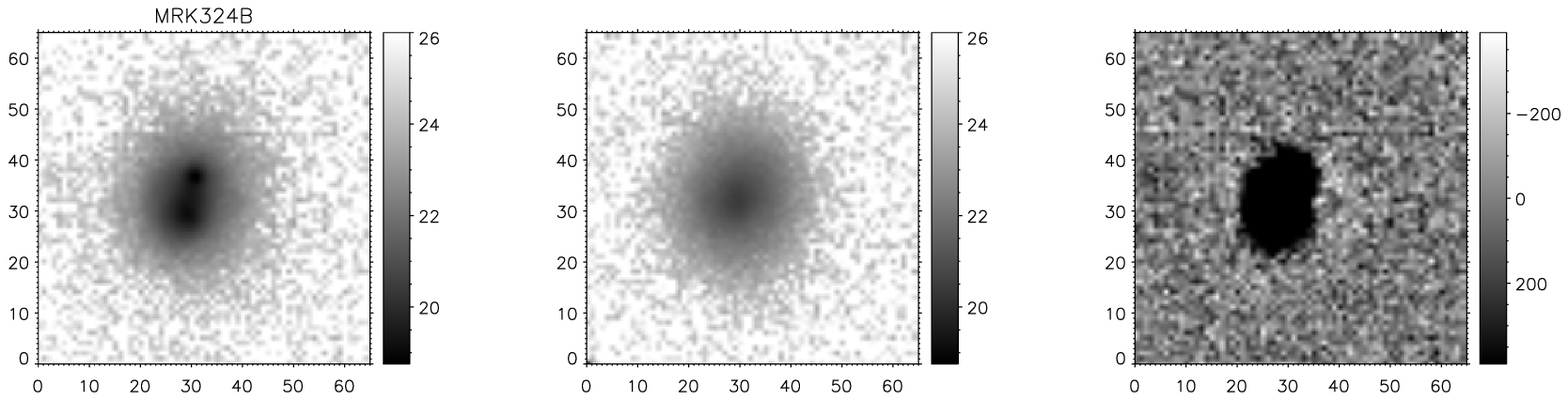}\\[7pt]
\caption[]{Atlas of galaxy images I: Galaxies with $Q=1$. Continued}
\end{figure*}

\begin{figure*}
\ContinuedFloat
\includegraphics[width=0.98\textwidth,angle=0]{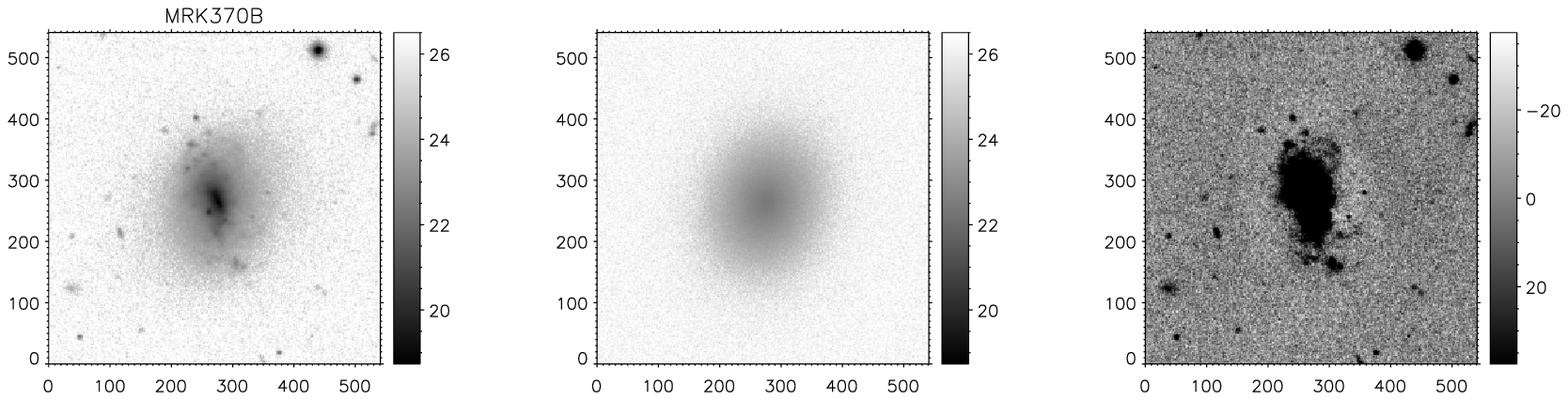}\\[7pt] 
\includegraphics[width=0.98\textwidth,angle=0]{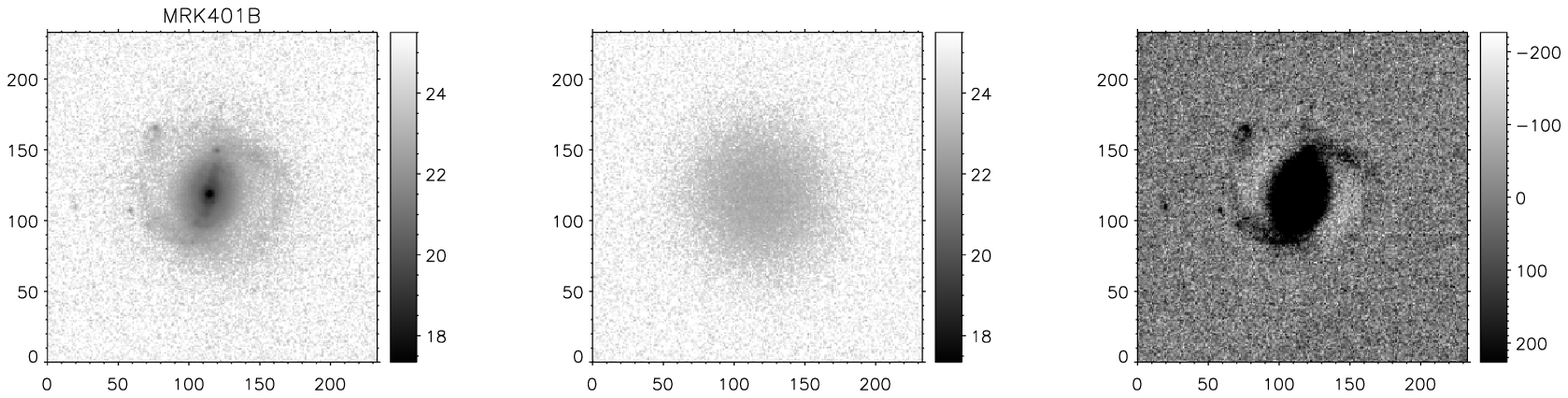}\\[7pt] 
\includegraphics[width=0.98\textwidth,angle=0]{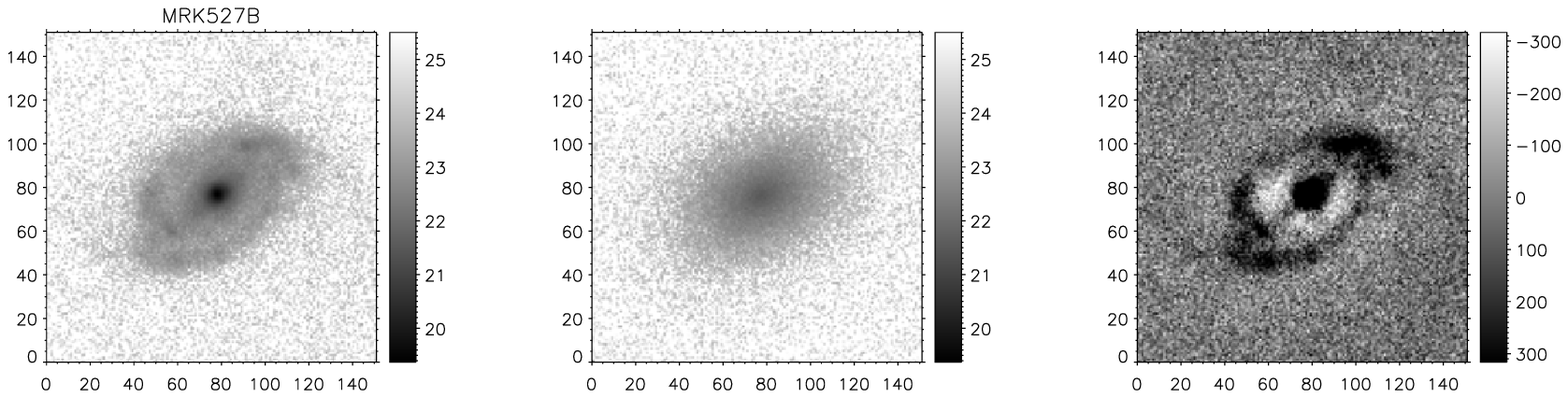}\\[7pt] 
\includegraphics[width=0.98\textwidth,angle=0]{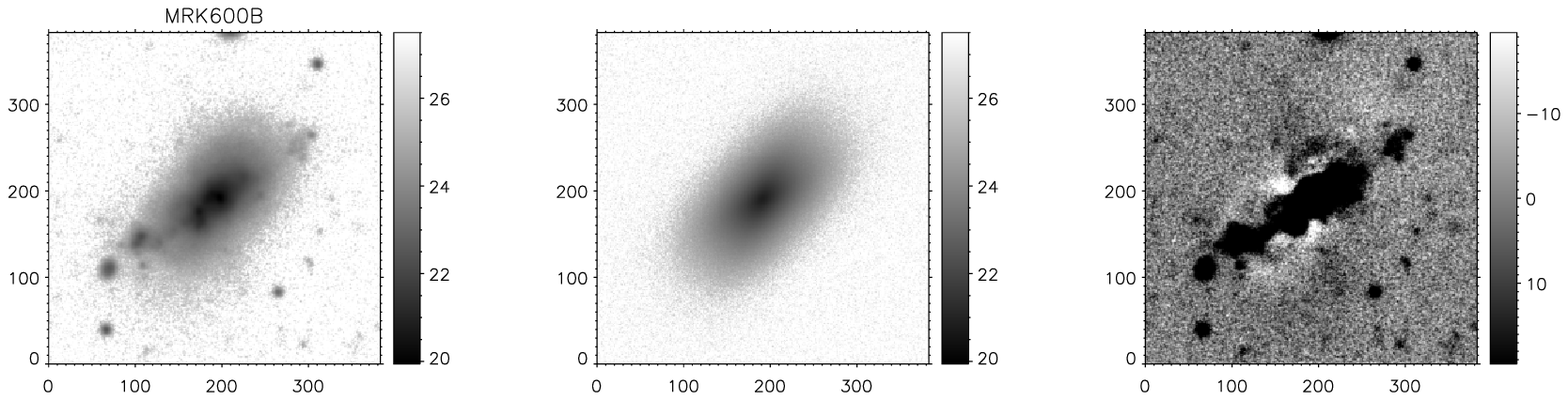}\\[7pt] 
\caption[]{Atlas of galaxy images I: Galaxies with $Q=1$. Continued}
\end{figure*}

\begin{figure*}
\ContinuedFloat
\includegraphics[width=0.98\textwidth,angle=0]{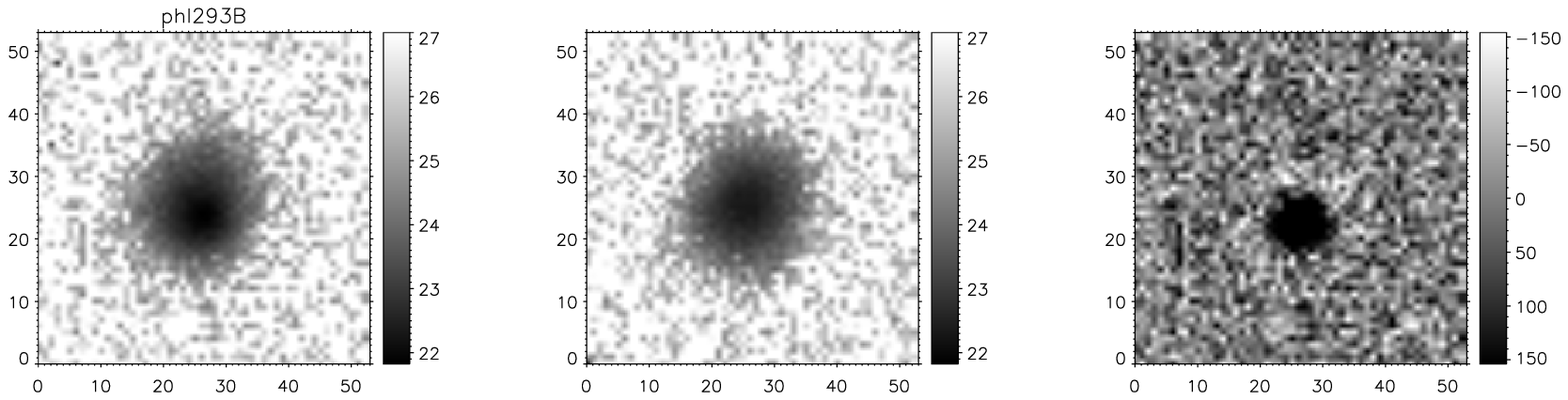}\\[7pt] 
\includegraphics[width=0.98\textwidth,angle=0]{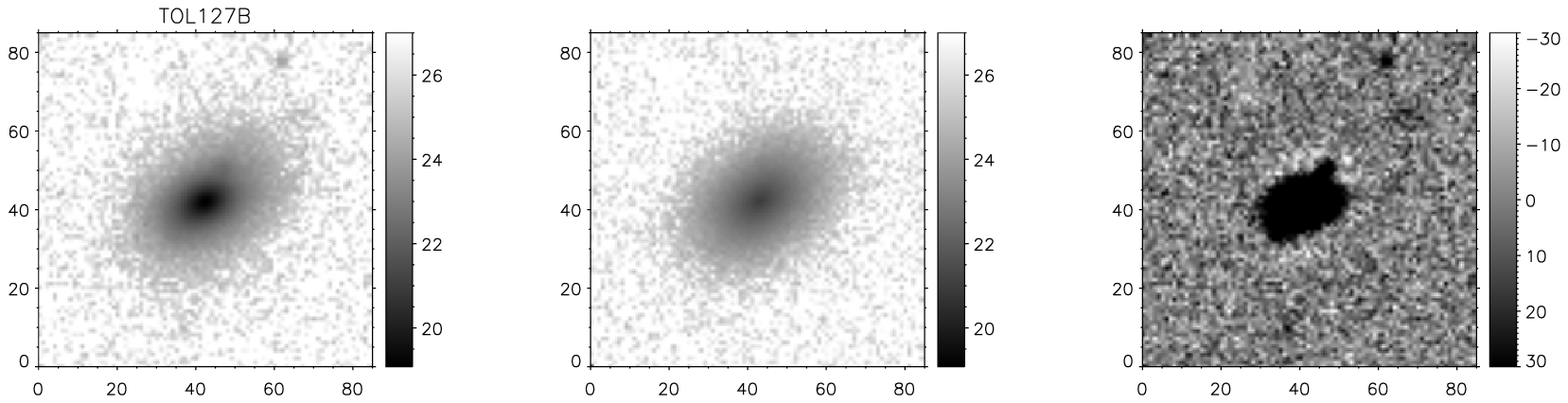}\\[7pt] 
\includegraphics[width=0.98\textwidth,angle=0]{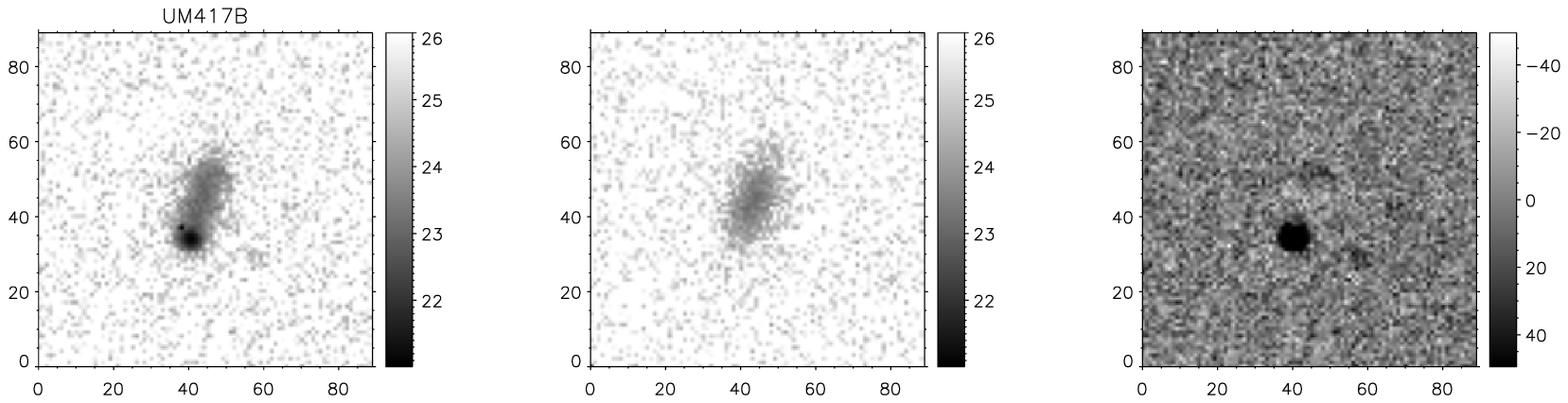}\\[7pt] 
\includegraphics[width=0.98\textwidth,angle=0]{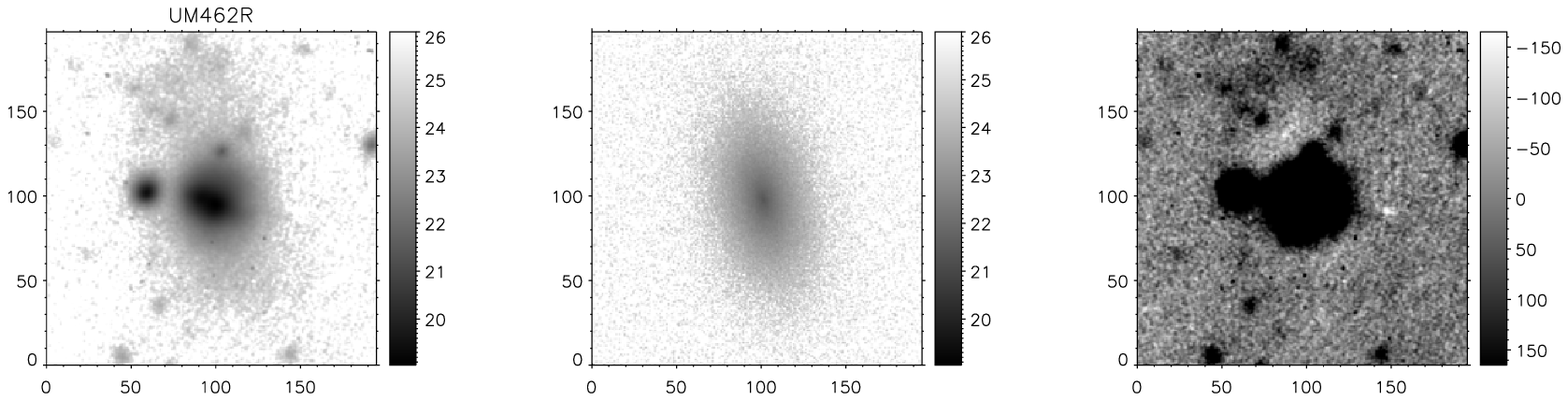}\\[7pt] 
\caption[]{Atlas of galaxy images I: Galaxies with $Q=1$. Continued}
\end{figure*}

\begin{figure*}
\ContinuedFloat
\includegraphics[width=0.98\textwidth,angle=0]{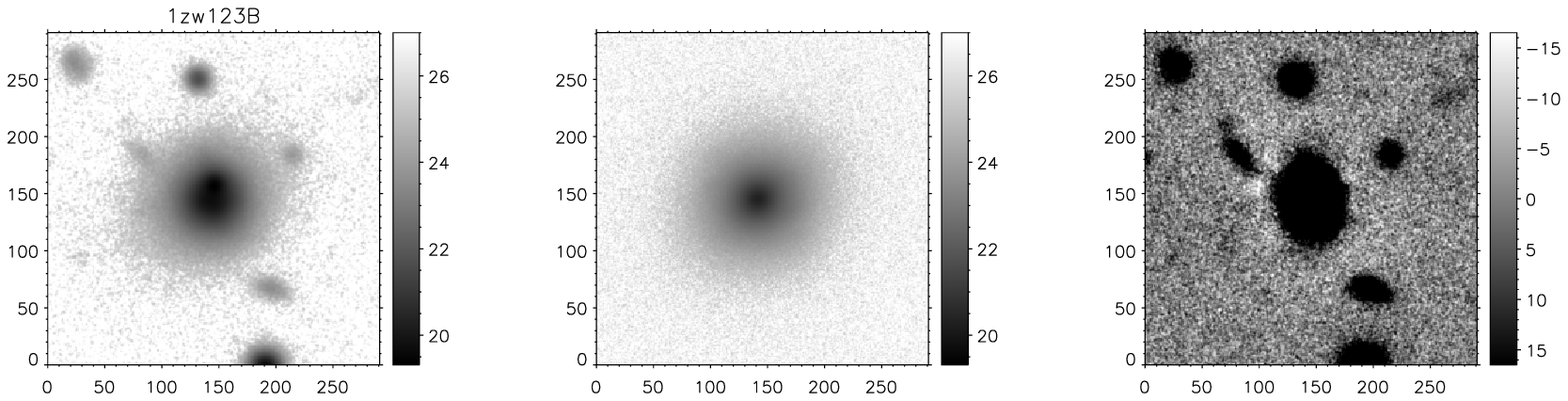}\\[7pt] 
\includegraphics[width=0.98\textwidth,angle=0]{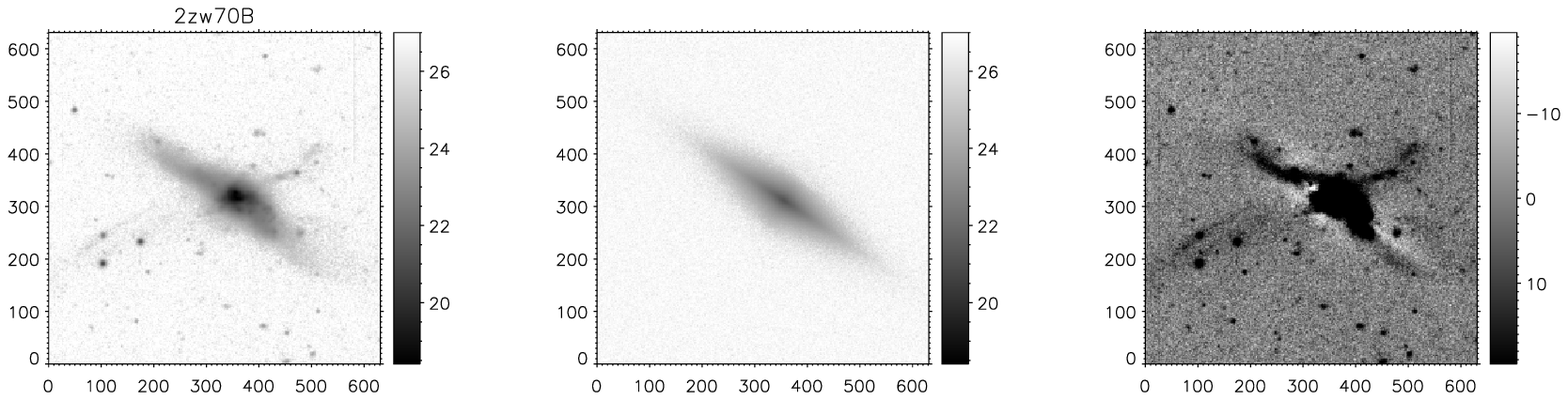}\\[7pt] 
\includegraphics[width=0.98\textwidth,angle=0]{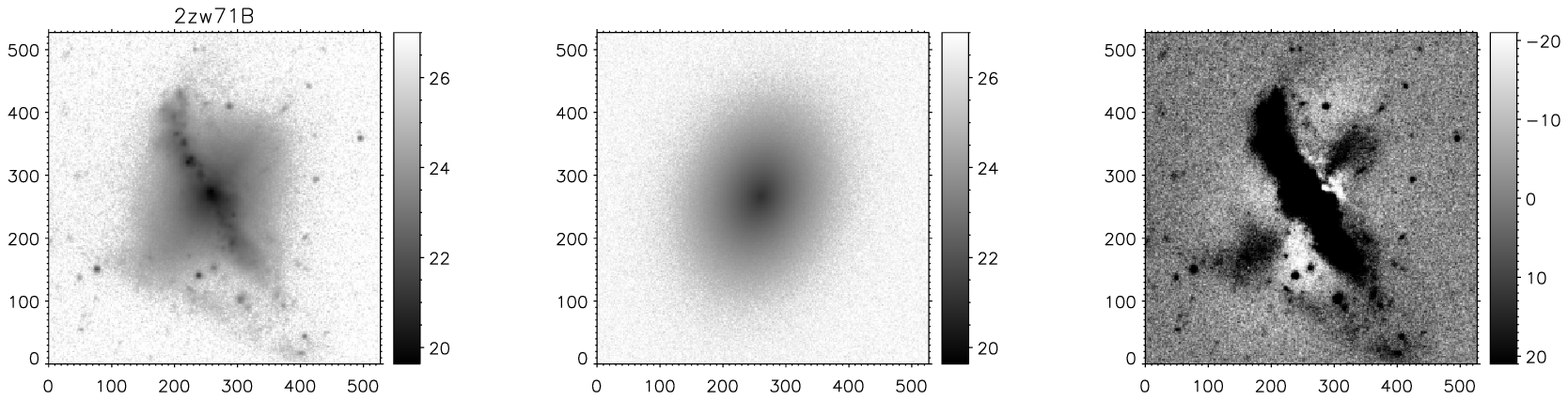}\\[7pt] 
\includegraphics[width=0.98\textwidth,angle=0]{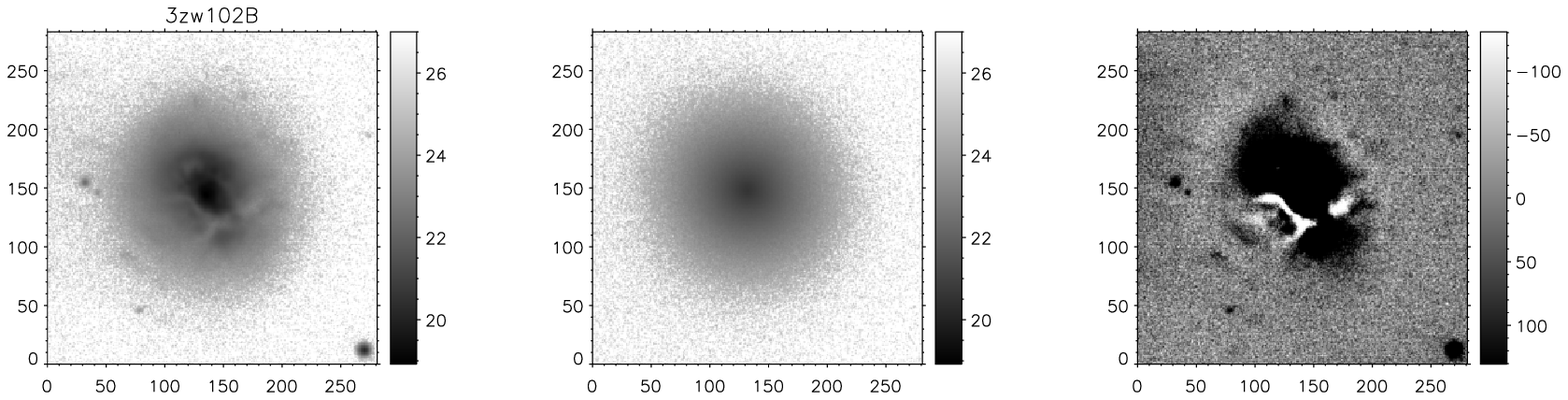}\\[7pt] 
\caption[]{Atlas of galaxy images I: Galaxies with $Q=1$. Continued}
\end{figure*}
\clearpage

\begin{figure*}
\includegraphics[width=0.98\textwidth,angle=0]{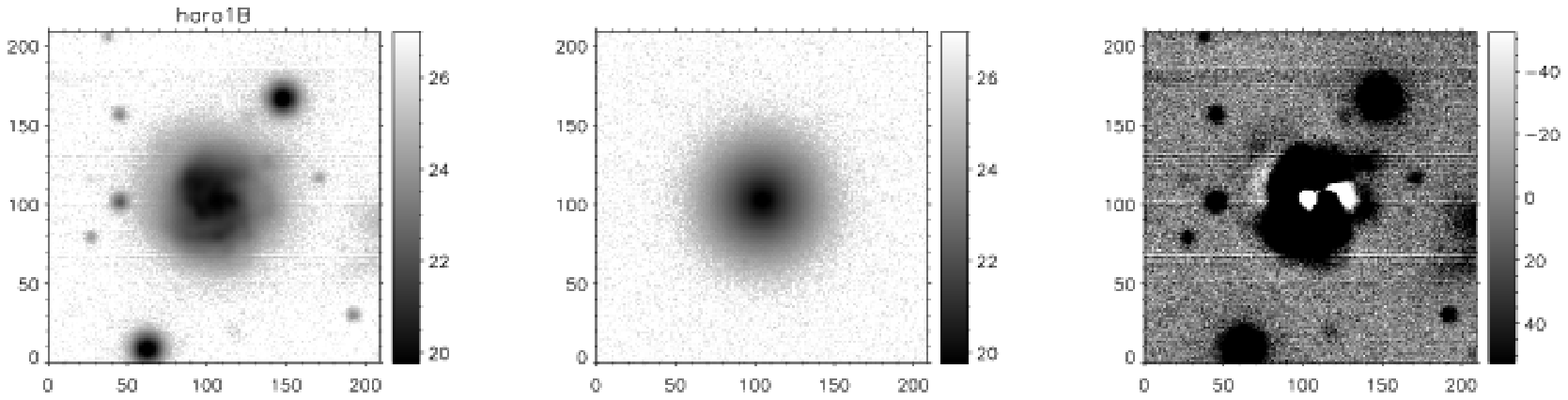}\\[7pt] 
\includegraphics[width=0.98\textwidth,angle=0]{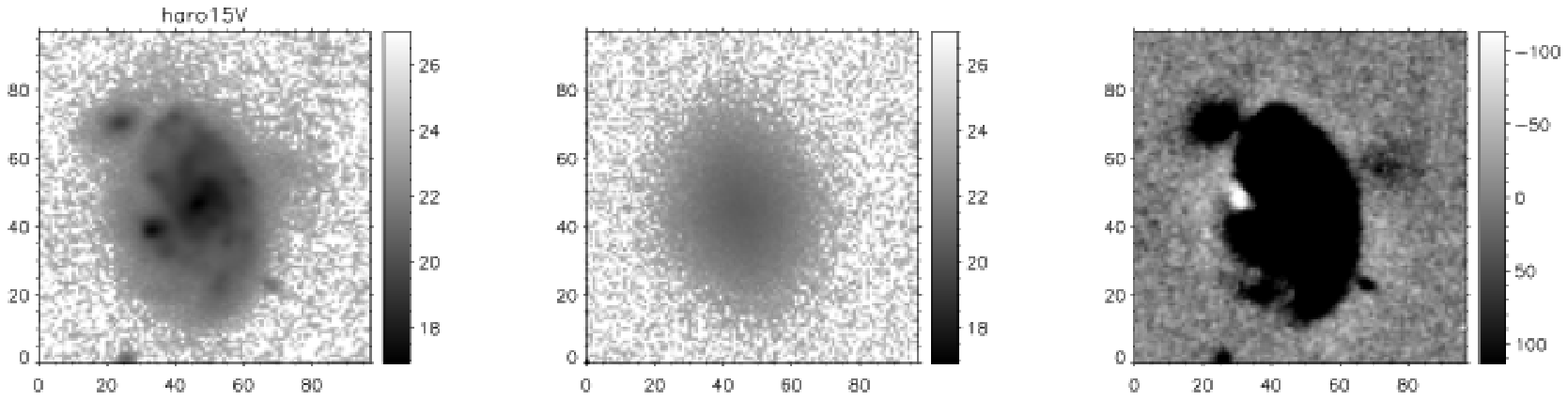}\\[7pt] 
\includegraphics[width=0.98\textwidth,angle=0]{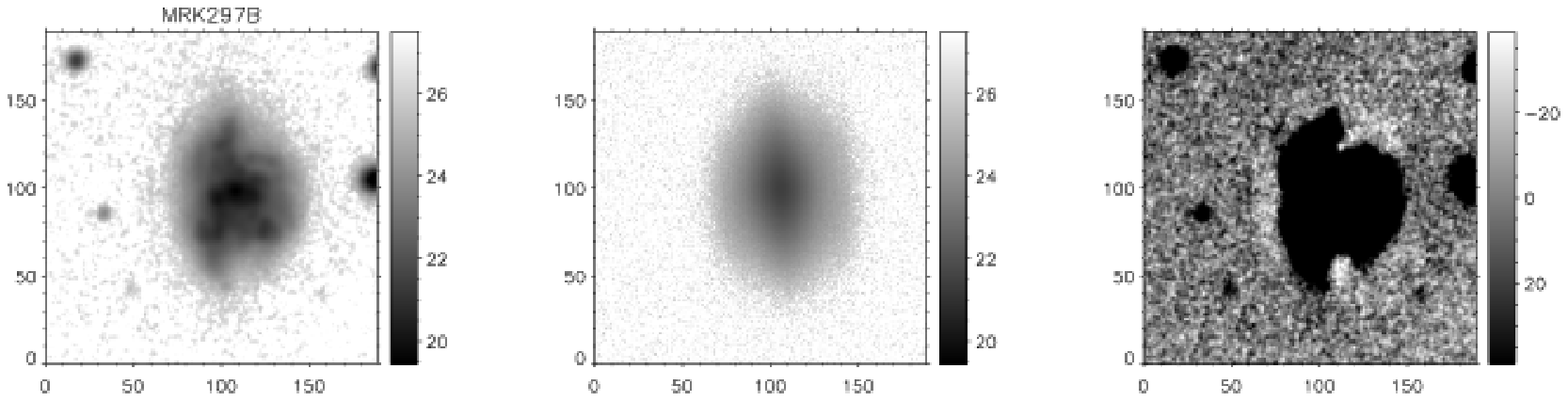}\\[7pt] 
\includegraphics[width=0.98\textwidth,angle=0]{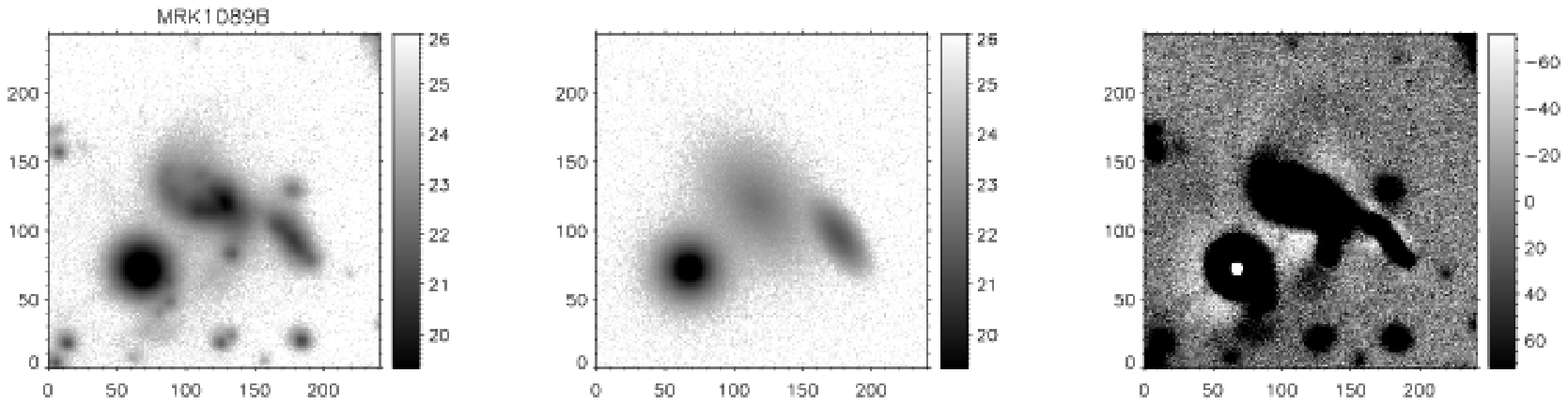}\\[7pt] 
\caption[]{Atlas of galaxy images II: Galaxies with $Q=2$. \footnotesize{Broad-band galaxy images (left), the best 2D \mbox{S\'ersic} model (centre) and the residual image (right) of the 8 galaxies fitted with quality index $Q=2$ in the full sample of 28 BCGs. The description is the same as in Fig.~\ref{Ap1}.}}
\end{figure*}

\begin{figure*}
\ContinuedFloat
\includegraphics[width=0.98\textwidth,angle=0]{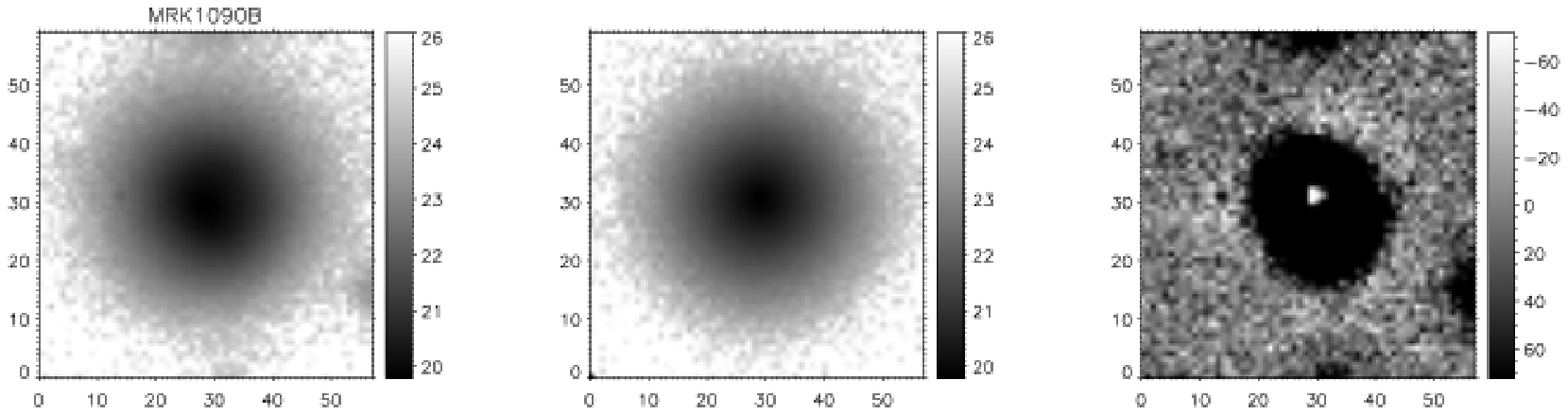}\\[7pt] 
\includegraphics[width=0.98\textwidth,angle=0]{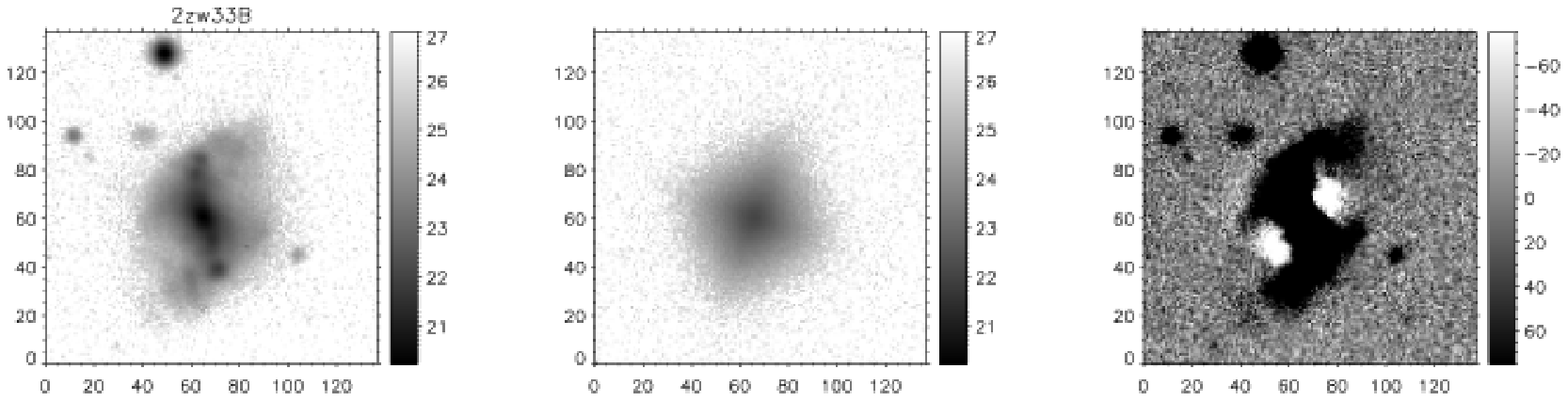}\\[7pt] 
\includegraphics[width=0.98\textwidth,angle=0]{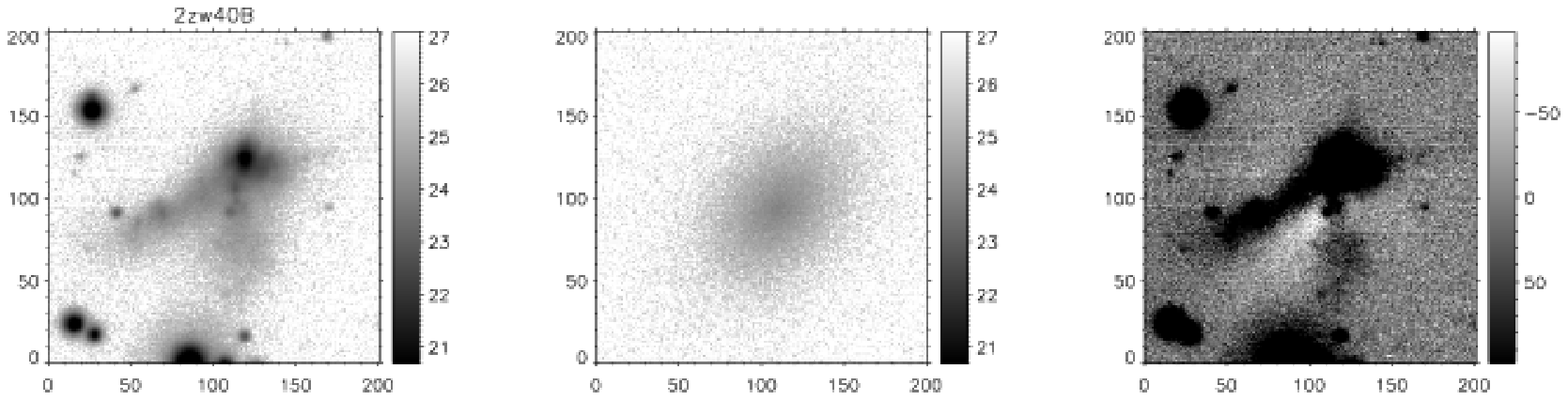}\\[7pt] 
\includegraphics[width=0.98\textwidth,angle=0]{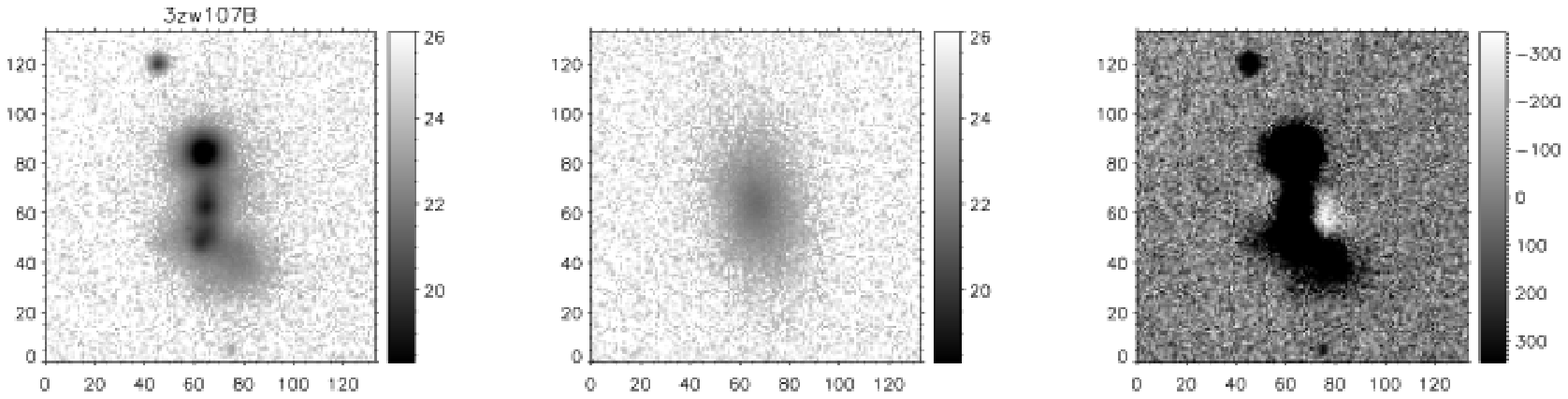}\\[7pt] 
\caption[]{\label{Ap2} Atlas of galaxy images II: Galaxies with $Q=2$. Continued}
\end{figure*}

\clearpage

\end{document}